\documentclass{aa}

\makeatletter
\renewcommand*\aa@pageof{, page \thepage{} of \pageref*{LastPage}}
\usepackage{natbib}
\bibpunct{(}{)}{;}{a}{}{,}
\usepackage{enumitem}
\usepackage{graphicx}
\usepackage{txfonts}
\usepackage[colorlinks=true,allcolors=blue]{hyperref}

\begin{document}

   \title{Near-field imaging of local interference in radio interferometric data}

   \subtitle{Impact on the redshifted 21 cm power spectrum}

\author{S. Munshi\inst{1} \and F. G. Mertens\inst{2,1} \and L. V. E. Koopmans\inst{1} \and M. Mevius\inst{3} \and A. R. Offringa\inst{3,1} \and B. Semelin\inst{2} \and C. Viou\inst{4} \and A. Bracco\inst{5,6} \and S. A. Brackenhoff\inst{1} \and E. Ceccotti\inst{1,7} \and J. K. Chege\inst{1,3} \and A. Fialkov\inst{8,9} \and L. Y. Gao\inst{10,1} \and R. Ghara\inst{11} \and S. Ghosh\inst{1} \and A. K. Shaw\inst{12} \and P. Zarka\inst{13,4} \and S. Zaroubi\inst{14,1} \and B. Cecconi\inst{13,4} \and S. Corbel\inst{15} \and J. N. Girard\inst{13} \and J. M. Grie{\ss}meier\inst{16,4} \and O. Konovalenko\inst{17} \and A. Loh\inst{13,4} \and P. Tokarsky\inst{17} \and O. Ulyanov\inst{17} \and V. Zakharenko\inst{17}}
    
    \institute{Kapteyn Astronomical Institute, University of Groningen, P.O. Box 800, 9700 AV Groningen, The Netherlands \and LUX, Observatoire de Paris, PSL Research University, CNRS, Sorbonne Université, F-75014 Paris, France \and ASTRON, PO Box 2, 7990 AA Dwingeloo, The Netherlands \and ORN, Observatoire de Paris, Université PSL, Univ Orléans, CNRS, 18330 Nançay, France \and INAF -- Osservatorio Astrofisico di Arcetri, Largo E. Fermi 5, 50125 Firenze, Italy \and Laboratoire de Physique de l'Ecole Normale Sup\'erieure, ENS, Universit\'e PSL, CNRS, Sorbonne Universit\'e, Universit\'e de Paris, F-75005 Paris, France \and INAF -- Istituto di Radioastronomia, Via P.~Gobetti 101, 40129 Bologna, Italy \and Kavli Institute for Cosmology, Madingley Road, Cambridge CB3 0HA, UK \and Institute of Astronomy, University of Cambridge, Madingley Road, Cambridge CB3 0HA, UK \and Liaoning Key Laboratory of Cosmology and Astrophysics, College of Sciences, Northeastern University, Shenyang 110819, China \and Department of Physical Sciences, Indian Institute of Science Education and Research Kolkata, Mohanpur, WB 741 246, India \and Department of Computer Science, University of Nevada Las Vegas, 4505 S. Maryland Pkwy., Las Vegas, NV 89154, USA \and LIRA, Observatoire de Paris, Université PSL, CNRS, Sorbonne Université, Université Paris Cité, 5 place Jules Janssen, 92195 Meudon, France \and ARCO (Astrophysics Research Center), Department of Natural Sciences, The Open University of Israel, 1 University Road, PO Box 808, Ra’anana 4353701, Israel \and Université Paris Cité and Université Paris Saclay, CEA, CNRS, AIM, 91190 Gif-sur-Yvette, France \and LPC2E, OSUC, Univ Orleans, CNRS, CNES, Observatoire de Paris, F-45071 Orleans, France \and Institute of Radio Astronomy, Mystetstv St. 4, 61002, Kharkiv, Ukraine}

   \date{Received --; accepted --}
 
  \abstract
    {Radio-frequency interference (RFI) is a major systematic limitation in radio astronomy, particularly for science cases requiring high sensitivity, such as 21 cm cosmology. Traditionally, RFI is dealt with by identifying its signature in the dynamic spectra of visibility data and flagging strongly affected regions. However, for RFI sources that do not occupy narrow regions in the time-frequency space, such as persistent local RFI, modeling these sources could be essential to mitigating their impact. This paper introduces two methods for detecting and characterizing local RFI sources from radio interferometric visibilities: matched filtering and maximum a posteriori (MAP) imaging. These algorithms use the spherical wave equation to construct three-dimensional near-field image cubes of RFI intensity from the visibilities. The matched filter algorithm can generate normalized maps by cross-correlating the expected contributions from RFI sources with the observed visibilities, while the MAP method performs a regularized inversion of the visibility equation in the near field to construct image cubes in physical units as a function of frequency. We developed a full polarization simulation framework for RFI and demonstrated the methods on simulated observations of local RFI sources. The stability, speed, and errors introduced by these algorithms were investigated, and, as a demonstration, the algorithms were applied to a subset of NenuFAR observations to perform spatial, spectral, and temporal characterization of two local RFI sources. We used simulations to assess the impact of local RFI on images, the $u\varv$ plane, and cylindrical power spectra, and to quantify the level of bias introduced by the algorithms in order to understand their implications for the estimated 21 cm power spectrum with radio interferometers. The near-field imaging and simulation codes are publicly available in the Python library \texttt{nfis}.}


   \maketitle

\defcitealias{munshi2024first}{M24}

\section{Introduction}\label{sec:intro}
The mitigation of radio frequency interference (RFI) is a persistent and growing challenge in radio astronomy. With the emergence of the next generation of telescopes with wider bandwidths and increased sensitivities, the overlap of observing spectral windows with bands affected by interference becomes inevitable, and the need to address low-level RFI in radio observations becomes even more crucial. The simultaneous technological advancement of the communication industry brings with it increased contamination from terrestrial transmitters, airplane communications \citep{gehlot2024transient}, and swarms of satellites in low-Earth orbits (\citealt{di2023unintended}; \citealt{grigg2023detection}; \citealt{bassa2024bright}; \citealt{grigg2025enhanced}; Zhang et al. under review). The increasing density of RFI sources in temporal, spectral, and spatial domains makes it necessary to develop and refine robust RFI mitigation strategies to preserve the maximum possible data integrity and enable science cases requiring high-sensitivity measurements.

The growth of radio interferometry has been accompanied by simultaneous progress in RFI mitigation techniques. While pre-correlation mitigation approaches, such as using filters in the front end targeting specific frequency bands and flagging raw voltage streams at high time resolutions, are sometimes essential in harsh RFI environments \citep{baan2004radio,niamsuwan2005examination}, post-correlation RFI detection and flagging are still almost always necessary to improve data quality. These methods primarily use RFI identification techniques in time-frequency space followed by thresholding and flagging of data points identified to be affected by RFI, both in imaging and beamformed observations. This process can be performed manually by inspecting dynamic spectra per baseline \citep{lane2005postcorrelation}, but automated detection and flagging algorithms have now become standard practice \citep{middelberg2006automated,offringa2010post,offringa2012morphological}. Alternative statistical techniques for RFI detection use deviations from the expected exponential distribution of the power spectral density to identify RFI-affected frequency channels \citep{fridman2001rfi,deshpande2005correlations,nita2007radio,gary2010wideband}. Recently, there has been increasing interest in machine learning methods that are trained to recognize complex RFI patterns and automate flagging \citep{wolfaardt2016machine,akeret2017radio,mesarcik2022learning}.

These RFI flagging techniques are optimal when the interference occupies narrow volumes in time-frequency-baseline space. Beyond simple detection and flagging, subtraction approaches have been developed to isolate and remove RFI contribution. For example, spatial filtering and subspace projection techniques identify and null directions of the RFI source through beamforming and decompose the data into orthogonal components, enabling the RFI to be isolated in one or more principal components, which can be subtracted while preserving the astronomical signals \citep{leshem2000multichannel, ellingson2002subspace,kocz2010radio}. While effective for identifying and subtracting persistent strong RFI, these approaches have the potential to introduce a bias in the measurement of the signal of interest. Modeling and subtraction is an alternative approach to RFI removal, where detailed characterization of the RFI source is performed and subtracted from the visibility data. This is particularly relevant for observatories suffering from persistent RFI sources either near the array or from satellites in deterministic trajectories. There have been several efforts to demonstrate the subtraction of RFI utilizing the stationarity of ground-based sources compared to the sky signal \citep{perley2003evla,cornwell2004rfi}, which is difficult to implement for phase centers located close to the celestial poles where even sky sources are relatively stationary with respect to the array. Several approaches to image in the near field have been explored by \cite{carter1988refocusing}, \cite{cornwell2004correction}, \cite{cornwell2004rfi}, \cite{lazio2009near}, and \cite{prabu2023near} by performing a near-field refocusing of the far-field equations. While these approaches require a priori knowledge of the distance to the emitters and are ideally suited for characterizing satellites in known trajectories, there have been demonstrations of algorithms that can infer the distance to the emitters from the data \citep{hu2023novel,ducharme2025altitude}. RFI localization algorithms through triangulation have been used extensively in remote sensing, and such an algorithm was used in the context of 21 cm cosmology with the Giant Metrewave Radio Telescope by \cite{paciga2011gmrt}. Recently, a Bayesian approach to jointly model calibration parameters and trajectories of satellite RFI in the near field of interferometers has been developed and demonstrated by \cite{finlay2023trajectory,finlay2025tabascal}.

The New Extension in Nan\c cay Upgrading loFAR \citep[NenuFAR:][]{zarka2012lss,zarka2015nenufar,zarka2020low} is a low-frequency radio interferometer located at the Nan\c cay radio observatory in France, that aims to detect the redshifted 21 cm signal from neutral hydrogen during cosmic dawn, the epoch when the first stars in the Universe formed \citep{mertens2021exploring}. The main challenges in 21 cm cosmology analyses are the orders of magnitude brighter foregrounds that obscure the faint background signal and additional systematics that prevent the thermal noise sensitivity of the instrument from being reached. This imposes stringent calibration requirements and the need to address extremely low-level RFI. Notable approaches to mitigate low-level systematics in 21 cm cosmology analyses include algorithms such as Sky-Subtracted Incoherent Noise Spectra \citep[SSINS;][]{wilensky2019absolving,wilensky2023evidence}, which can mitigate RFI below single baseline noise levels, and approaches to mitigate instrumental coupling between feeds through fringe rate filtering \citep{kern2019mitigating,kern2020mitigating,charles2023use,charles2024mitigating,garsden2024demonstration}. The first analysis of NenuFAR data in the context of 21 cm cosmology \citep[][hereafter \citetalias{munshi2024first}]{munshi2024first} identified that local RFI sources near the core of the array contribute significantly to the residuals in the data after foreground removal. In this paper, we develop techniques to perform realistic near-field RFI simulations and spatial, spectral, and temporal characterization of local RFI sources. We demonstrate these techniques by characterizing the local RFI sources in NenuFAR data and assessing the impact of the RFI sources through simulations on images, the $u\varv$ plane, and 21 cm power spectra. In a follow-up paper, these methods will be applied to more data to assess their impact on improving the 21 cm power spectrum limits derived with NenuFAR.

The paper is organized as follows. In Sect. \ref{sec:response}, we describe the near-field response of an interferometer. In Sect \ref{sec:nfi}, we introduce the near-field imaging techniques and demonstrate them on simulated radio interferometric data. In Sect. \ref{sec:rfi_nenufar}, we apply the methods to a subset of NenuFAR observations to perform spectral and temporal characterization of two local RFI sources. In Sect. \ref{sec:far_field}, we use near-field simulations to understand the impact of local RFI sources on far-field data such as the power spectrum, $u\varv$ plane, and images. In Sect. \ref{sec:discussion}, we discuss the strengths and limitations of the algorithms and future prospects.

\section{Array response to near-field RFI sources}\label{sec:response}
The boundary between the near and far fields for an instrument of dimension $D$ observing at a wavelength $\lambda$ is typically defined by the Fraunhofer distance ($d_{\mathrm{F}}$) given by $d_{\mathrm{F}} = 2D^2/\lambda$. In this section, we derive the far- and near-field visibility equations. While the latter is valid for nearly all RFI sources, even those in low-Earth orbits, the far-field visibility equation is used to assess the impact of the presence of near-field RFI emission on traditional far-field images and the 21 cm power spectrum when assuming that all emission comes from the far field.

\subsection{Far-field visibilities}
Astronomical sources lie in the far field of an interferometer, and the wavefront from these sources can be approximated as a plane wave. This is the basis of standard far-field interferometric imaging, where the delay in the signals arriving at the two stations constituting a baseline is proportional to the dot product of the baseline vector ($\mathbf{b}$) and the source (unit) vector ($\hat{\mathbf{s}}$) at frequency $\nu$. The spatial coherence or visibility function corresponding to a sky brightness matrix, $\mathbf{I}(\hat{\mathbf{s}},\nu)$, measured by a baseline, after applying geometric delay correction to a phase center ($\hat{\mathbf{p}}$), can then be written as \citep{hamaker1996understanding,smirnov2011revisiting,thompson2017interferometry}
\begin{equation}\label{eq:far_field}
\mathbf{V}(\nu)=\mathbf{G}_{p}\left(\int \mathbf{E}_{p}\mathbf{I}\mathbf{E}_{q}^{H} \, \mathrm{exp}\left[-\frac{2\pi \mathrm{i} \nu}{c}\left(\mathbf{b}\cdot\left(\hat{\mathbf{s}} - \hat{\mathbf{p}}\right)\right)\right] \mathrm{d} \Omega\right)\mathbf{G}_{q}^{H}.
\end{equation}
Here parameters in uppercase boldface are Jones matrices; d$\Omega$ is the differential solid angle on the unit sphere; $\mathbf{G}_{p}(\nu)$ and $\mathbf{E}_{p}(\hat{\mathbf{s}},\nu)$ are the direction-independent (DI) and direction-dependent (DD) gains, respectively, for the $p$-th station; and the superscript $H$ indicates a Hermitian transpose. The DI gains are corrected in the visibility data through calibration against a known sky model. The main contributor toward the DD gains is the instrumental primary beam, which is often, to first order, considered to have the same functional form for all stations for an array composed of stations with the same configuration. Then the term $\mathbf{E}_{p}\mathbf{I}\mathbf{E}_{q}^{H}$ can be approximated as an apparent intensity distribution seen by the array given by $\mathbf{I}'(\hat{\mathbf{s}},\nu) = \mathbf{E}\mathbf{I}\mathbf{E}^{H}$. Considering a three-dimensional (3D) coordinate system with the third axis pointing along the phase center, with the baseline coordinates given by $\mathbf{b}=(U,V,W)$, in physical units, and the source coordinates given by $\hat{\mathbf{s}} = (l,m,n=\sqrt{1-l^2-m^2})$, the visibility function reduces to
\small
\begin{equation}
\mathbf{V}(\nu)=\iint_{l,m} \frac{1}{n}\mathbf{I}'(l,m,n,\nu) \, \mathrm{exp}\left[-\frac{2\pi \mathrm{i} \nu}{c}\left(Ul+Vm+W(n-1)\right)\right] \mathrm{d}l\mathrm{d}m.
\end{equation}
\normalsize
For instruments with small fields of view where the flat sky approximation ($l,m << 1$) holds, this reduces to a two-dimensional (2D) Fourier relation between the visibilities in the $u\varv$ plane and the sky ($l,m$) plane given by
\begin{equation}
\mathbf{V}(u,\varv,\nu)=\iint_{l,m} \mathbf{I}'(l,m,\nu) \, \mathrm{exp}\left[-2 \pi i\left(ul+\varv m\right)\right] \mathrm{d}l\mathrm{d}m,
\end{equation}
where $u=U\nu/c$ and $\varv=V\nu/c$. 

\subsection{Near-field visibilities}
Most terrestrial RFI sources fall in the near-field regime of radio interferometers where a plane wave approximation is not valid. For example, even NenuFAR, which is an extreme case of a compact interferometer at low frequencies, has $d_{\mathrm{F}} >> 2000\,$km at $\nu=60\,$MHz, which means that satellites in low-Earth orbits would be in the near field of the instrument. The spatial dependence of the electric field at a location $\mathbf{r}$ due to an RFI emitter at $\mathbf{r}'$ can be described using the Green's function $G(\mathbf{r},\mathbf{r}')$ corresponding to the 3D inhomogeneous Helmholtz equation with a delta function source term \citep[e.g.,][]{colton1998inverse}. For each frequency, the Green's function in free space is a spherical wave of the form
\begin{equation}
G_{\nu}\left(\mathbf{r}, \mathbf{r}'\right)=\frac{\mathrm{exp}\left({\frac{2\pi \mathrm{i} \nu}{c}|\mathbf{r}-\mathbf{r}'|}\right)}{|\mathbf{r}-\mathbf{r}'|} .
\end{equation}
Consider a field of emitters with spectral power density distribution (in units of Watt Hz$^{-1}$m$^{-3}$) given by $P_d(\mathbf{r}',\nu)$. The DI calibrated visibility measured on a baseline $\mathbf{b}$ formed by two stations located at $\mathbf{r}_p$ and $\mathbf{r}_q$ after geometric phasing to the phase center $\hat{\mathbf{p}}$ is obtained by cross-correlating the electric fields received by the $p$-th and $q$-th elements, and is given by
\small
\begin{equation}
\mathbf{V}_{pq}(\nu) = \int\frac{\mathbf{E}_{p}\mathbf{E}_{q}^{H}P_d(\mathbf{r}',\nu)\mathrm{exp}\left[-\frac{2\pi \mathrm{i}\nu}{c}\left(|\mathbf{r}_p - \mathbf{r}'|-|\mathbf{r}_q - \mathbf{r}'|-\mathbf{b}\cdot\hat{\mathbf{p}}\right)\right]}{|\mathbf{r}_p - \mathbf{r}'||\mathbf{r}_q - \mathbf{r}'|}\mathrm{d}^{3}\mathbf{r}'.
\end{equation}
\normalsize
In the near field, $\mathbf{E}_{p}$ depends on the direction of $\mathbf{r}'-\mathbf{r}_p$ and the assumption $\mathbf{E}_{p}$ = $\mathbf{E}_{q}$ does not hold in general even for identical stations. Here, the RFI emitters are assumed to be unpolarized and isotropic. Additionally, it has been assumed that the electric field propagation near the plane of the array is not affected by the array itself. Though all these assumptions are likely to break down in reality, the techniques developed in this paper using these assumptions work well for NenuFAR data as shown later in the paper.\footnote{This is possibly because, even if the emitters are linearly polarized with a vertical component, most receiving dipoles lie along the ground plane and receive the horizontal component parallel to the ground. The $P(\mathbf{r}',\nu)$ then corresponds to the horizontal component of the incoming radiation.} Both polarization and propagation effects could, in principle, be included in the formalism, but this is beyond the scope of the current paper. Under the current assumptions, the measured visibility coherence matrix gets its polarization state solely due to instrumental polarization. Assuming a distribution of $N$ isotropic emitters with spectral powers given by $P(\mathbf{r}_i,\nu)$, the equation can be discretized to
\begin{equation}\label{eq:near_field}
\mathbf{V}_{pq}(\nu)=\sum_{i=1}^{N} \dfrac{\mathbf{E}_{pi}\mathbf{E}_{qi}^{H}P(\mathbf{r}_i,\nu)}{d_{p}(\mathbf{r}_i)d_{q}(\mathbf{r}_i)}\, \mathrm{exp}\left[-\frac{2\pi \mathrm{i} \nu}{c} \left(d_{p}(\mathbf{r}_i)-d_{q}(\mathbf{r}_i)-\mathbf{b}\cdot\hat{\mathbf{n}}\right)\right].
\end{equation}
Here $d_{p}(\mathbf{r}_i)=|\mathbf{r}_i-\mathbf{r}_p|$ is the distance between the RFI source located at $\mathbf{r}_i$ and the $p$-th interferometric element located at $\mathbf{r}_p$. The geometric delay is proportional to the physical path difference given by $d_{p}(\mathbf{r}_i)-d_{q}(\mathbf{r}_i)$, and the spherical wave propagation induces a free space attenuation of the received flux corresponding to an inverse square law. Equation (\ref{eq:near_field}) cannot be simplified to a 2D Fourier relation, since the phase cannot be cast in the form of a dot product between spatial locations of the RFI sources ($\mathbf{r}_i$) and a combination of the station coordinates $\mathbf{r}_p$ and $\mathbf{r}_q$. Throughout most of the remainder of this paper, we assume that Eq. (\ref{eq:near_field}) describes the visibilities of near-field RFI sources under the conditions stated above (i.e., isotropic unpolarized emitters and no propagation effects).

\section{Near-field imaging}\label{sec:nfi}
In this section, we present methods for generating maps of local RFI sources from radio interferometric visibilities, using the spherical wave propagation equations described previously. We explore two alternative algorithms for constructing near-field images, each with distinct advantages and trade-offs in terms of performance in low signal-to-noise ratio (S/N) conditions, model accuracy, and computational efficiency.

\subsection{Simulations}
To demonstrate and assess the performance of the two near-field imaging methods, we performed simulations of local RFI sources in the context of NenuFAR. The visibility contributions from local RFI sources can be simulated on radio interferometric measurement sets using Eq. (\ref{eq:near_field}). The exact response of a baseline to the RFI source will depend on the radiation patterns and orientations of the individual dipoles measuring the X and Y polarizations. Let R($\mathbf{l}_a, \theta, \nu$) be the radiation pattern for a dipole where $\mathbf{l}_a$ is the dipole vector for the $a^{th}$ feed and $\theta$ is the angle between the vectors $\mathbf{l}_a$ and $\mathbf{r}'-\mathbf{r}_p$. For simplicity, in our simulations, we assume unpolarized isotropic RFI emitters and the receiving antennas to be composed of infinitesimal dipoles with a $\sin\theta$ radiation pattern. The DD Jones matrix for the $p^{th}$ station is then given by
\begin{equation}
\mathbf{E}_{pi} = \begin{bmatrix}
R(\mathbf{l}_x, \theta_{pxi}, \nu) & 0\\
0 & R(\mathbf{l}_y, \theta_{pyi}, \nu)
\end{bmatrix} = \begin{bmatrix}
\sin(\theta_{pxi}) & 0 \\
0 & \sin(\theta_{pyi})
\end{bmatrix}.
\end{equation}
Using this configuration, visibilities for synthesis observations of the north celestial pole (NCP) field with the NenuFAR station configuration were simulated on existing measurement sets. We note that the attenuation of RFI flux due to time and frequency smearing effects has not been considered in our simulations.

\subsection{Matched filter imaging}\label{sec:matched_filter}
Although Eq. (\ref{eq:near_field}) cannot be cast into a direct 2D Fourier transform equation, the distribution of RFI sources located on a grid in the near field of the instrument can be identified by comparing their expected contribution with the observed visibilities. This matched filter approach was first demonstrated on NenuFAR data by \cite{smeenk2020radio} and is further developed here using a mathematical framework and simulations to examine its implications. This method essentially produces spatial dirty image cubes in the near field of the array. 

We consider a 3D grid of size $N_x\times N_y\times N_z = N_g$ with coordinates ($x,y,z$) in physical space relative to the array. It should be noted that, ideally, the grid cell size must be of the order of $\lambda/2$ (= 2.5 m at 60 MHz) or smaller to make sure that residual phase due to a source away from a grid point does not decorrelate the signal. However, for phased arrays, the individual stations are typically much larger than $\lambda$, and the antennas placed across the extent of stations are phased toward the pointing direction, and their voltages are added before correlation, resulting in loss of spatial information at scales smaller than the station sizes. In the case of NenuFAR, the stations are $22\,$m across, but we consider the centroid of an individual station as its location. Thus, not accounting for the phase at each antenna ignores the array factor for both stations of the baseline under consideration, leading to errors in the calculation of the phase for a given baseline. We incorporate this effect into a baseline and source-dependent gain term $G_{pq}(x,y,z)$ with $|G_{pq}(x,y,z)|=1$ and $\langle G_{pq}(x,y,z)\rangle_{\mathrm{bl}}\rightarrow 0$, where $\langle\dots\rangle_{\mathrm{bl}}$ denotes an average over baselines. This effect is discussed in more detail in Sect. \ref{sec:matched_filter_alt} and Appendix \ref{sec:matched_filter_alt_data}, where we demonstrate the impact of these gains on simulated and observed data. For brevity, we denote the distance terms $d_p(x_i,y_i,z_i)$ and the gain terms $G_{pq}(x_i,y_i,z_i)$ as $d_{pi}$ and $G_{pqi}$ respectively, reducing Eq. ($\ref{eq:near_field}$) to
\begin{align}\label{eq:near_field_discrete}
\mathbf{V}_{pq}(\nu,t) = \sum_{i=1}^{N_g} 
&\frac{\mathbf{E}_{pi} \mathbf{E}_{qi}^{H} P(x_i,y_i,z_i,\nu)}{d_{pi} d_{qi}} \,G_{pqi} \nonumber \\
&\quad \times \mathrm{exp}\Biggl[-\frac{2\pi \mathrm{i} \nu}{c} 
\bigl( d_{pi} - d_{qi} - W_{pq}(t) \bigr)\Biggr] .
\end{align}

A method to recover the near-field map through the matched filter approach consists of the following steps: 
\begin{enumerate}
    \item Apply the inverse of the geometric phase ($e^{-2\pi \mathrm{i}\nu W_{pq}(t)/c}$) to the observed visibilities, which effectively phases the data toward the zenith in the far field.
    \item Average these data in time to smear out contributions from astronomical sources except those near the celestial poles.
    \item For each point in the chosen grid, cross-correlate the expected near-field phase with the observed visibilities. This is the matched filter operation and ensures that only contributions from grid points with RFI sources add up coherently in the subsequent step.
    \item Average the phased data along the frequency, compute the absolute value and average along the baselines.
\end{enumerate}
The $W$ coordinate for an NCP phase center is practically fixed in time, thus avoiding decorrelating the RFI source while phasing due to time or frequency averaging. For other phase centers, the data need to be at sufficient time and frequency resolution to avoid smearing while phasing back to the zenith. This condition is usually met for low frequencies at which NenuFAR operates. The sequence of operations in step (4) is chosen due to the presence of the $G_{pqi}$ term, and the necessity of these steps is discussed in Sect. \ref{sec:matched_filter_alt}. Next, we describe how this sequence of steps can provide a spatial heatmap of local RFI sources.
\begin{figure*}
    \includegraphics[width=\columnwidth]{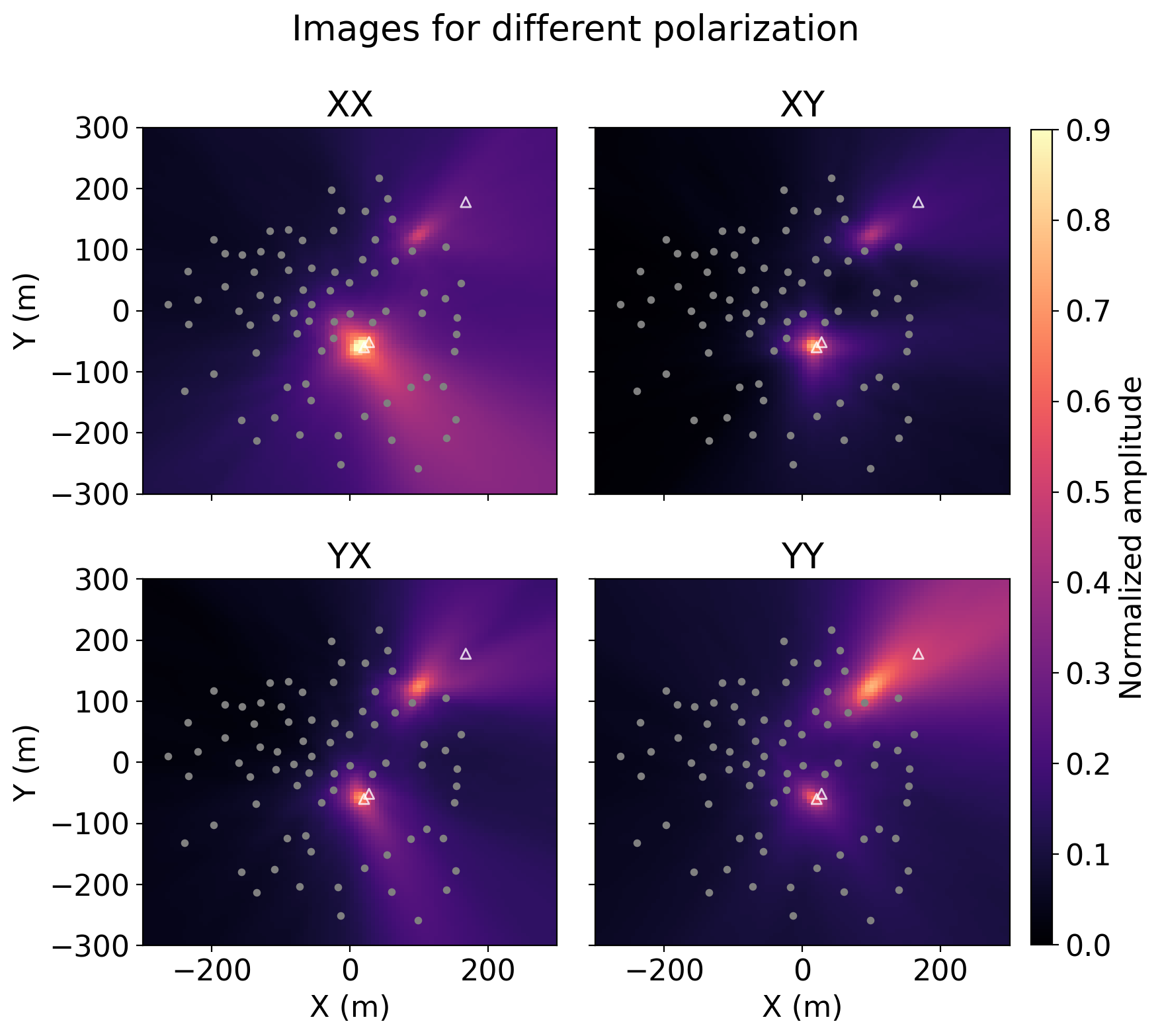}
    \includegraphics[width=\columnwidth]{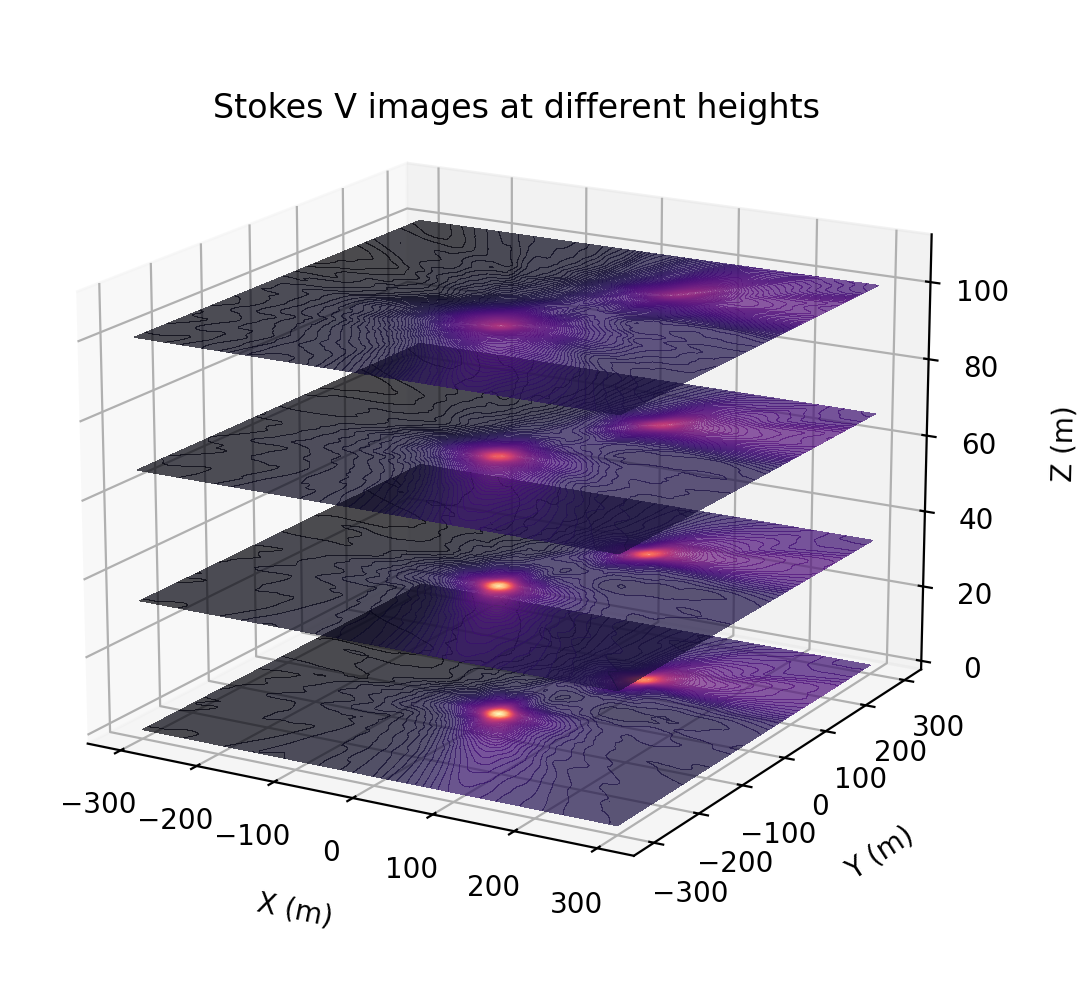}
    \caption{Near-field images of simulated datasets, constructed using the matched filter technique. The left panel shows the effect of instrumental polarization on the near-field images estimated per correlation. The gray dots indicate the locations of NenuFAR core stations, and the white hollow triangles indicate the locations of the three buildings within NenuFAR. The right panel shows images made at different heights above the ground.}
    \label{fig:mf_sim}
\end{figure*}
\subsubsection{Creating a 3D near-field dirty image cube}\label{sec:matched_filter_steps}
We consider two grid points: grid point $a$, where an RFI source is present, and grid point $b$, which represents any other grid point that is dominated by noise. The contributions from the visibilities at grid points $a$ and $b$ after steps 1, 2, and 3 are given by
\begin{align*}
    \mathbf{V}_{pq}^{a}(\nu) &= \frac{\mathbf{E}_{pa} \mathbf{E}_{qa}^{H}P(\nu)G_{pqa}}{d_{pa}d_{qa}},\nonumber\\
    \mathbf{V}_{pq}^{b}(\nu) &= \frac{\mathbf{E}_{pb} \mathbf{E}_{qb}^{H}P(\nu) G_{pqb}}{d_{pb}d_{qb}}\times\mathrm{exp}\left[\frac{-2\pi \mathrm{i} \nu}{c}\Delta d_{pqab}\right].
\end{align*}
Here $\Delta d_{pqab} = (d_{pa}-d_{pb})-(d_{qa}-d_{qb}) \neq 0$ in general. Averaging these contributions along frequency for both grid points gives
\begin{align*}
\langle \mathbf{V}_{pq}^{a}(\nu)\rangle_{\nu} &= \frac{\mathbf{E}_{pa} \mathbf{E}_{qa}^{H}\langle P(\nu)\rangle_{\nu} G_{pqa}}{d_{pa}d_{qa}}, \nonumber\\
\langle \mathbf{V}_{pq}^{b}(\nu)\rangle_{\nu} &= \frac{G_{pqb}\mathbf{E}_{pb} \mathbf{E}_{qb}^{H}}{d_{pb}d_{qb}}\times \Biggl< P(\nu) \mathrm{exp} \left[-\frac{2\pi \mathrm{i} \nu}{c}\Delta d_{pqab}\right]\Biggr>_{\nu} \approx 0.
\end{align*}
We note that here we assume that the DD Jones matrices vary slowly along the frequency. The subsequent steps of computing the absolute value and baseline averaging result in
\begin{align*}
\bigl<|\langle \mathbf{V}_{pq}^{a}(\nu)\rangle_{\nu}|\bigr>_{\mathrm{bl}} &= \langle P(\nu)\rangle_{\nu}\Biggl<\frac{|\mathbf{E}_{pa} \mathbf{E}_{qa}^{H}G_{pqa}|}{d_{pa}d_{qa}}\Biggr>_{\mathrm{bl}}, \nonumber\\
\bigl<|\langle \mathbf{V}_{pq}^{b}(\nu)\rangle_{\nu}|\bigr>_{\mathrm{bl}} &\approx 0.
\end{align*}
Thus, the effective power measured at a grid point that has an RFI source is the averaged spectral power over frequency multiplied by a factor that depends on the location of the grid point and all the baseline locations. So, while the maps produced in this way will, in general, have a non-zero effective power at the grid points near RFI sources, they will not have physical units that intrinsically describe the RFI source. Additionally, the frequency behavior is lost due to the necessity of performing the frequency averaging operation. We note that this way of constructing spatial near-field image cubes is analogous to far-field dirty images created by gridding and Fourier transforming visibilities. As in standard far-field imaging, the imaging done over a finite volume could miss RFI sources outside the reconstructed volume, although their sidelobes will leak into the cube. Unlike the finite unit sphere in far-field imaging, here, the volume that needs reconstruction should, in principle, be as large as the half-sphere volume of radius $d_{\mathrm{F}}$, outside which the RFI is in the far field. Reconstructing such large volumes in general is not needed and is currently also not feasible.

\subsubsection{Application to simulated data}\label{sec:mf_sim}
We used Eq. (\ref{eq:near_field}) to simulate visibilities to test the performance of the matched filter imaging. The dataset consists of two RFI sources located near the ground at a height defined by the average elevation of the NenuFAR core stations. The first source is located near the electronic containers within the NenuFAR core, and the other is located near the northeast of the core. This model is motivated by real 
NenuFAR observations. The image cubes were constructed on a $100\times 100\times 4$ grid in the $XYZ$ space, with grid resolutions of $6\,$m in $X$ and $Y$ directions, and $33\,$m in the $Z$ direction. This resolution was sufficient to sample the 3D point spread function (PSF) of the RFI sources.

The left panel of Fig.~\ref{fig:mf_sim} shows the effect of instrumental polarization in near-field images constructed using the matched filter approach. Here, the images were created separately for the different elements of the visibility coherence matrix. We see that the central source (hereafter source 1) has the highest amplitude in XX and the least in YY. The situation is reversed for the source in the northeast (hereafter source 2), which has higher amplitudes in YY polarization. The reason is that the X dipoles are oriented in the southwest to northeast direction, while the Y dipoles are oriented southeast to northwest. Since the majority of NenuFAR stations are located to the northwest of source 1, the reception patterns of the X dipoles are more sensitive to it, while the Y dipoles pick up source 2 more strongly. This is verified in Sect. \ref{sec:nf_data}, where images made from actual NenuFAR data are seen to reproduce these signatures. The right panel of Fig.~\ref{fig:mf_sim} shows the near-field images constructed at different heights above the ground. The images pick up the sources most strongly at $Z=0$ since the input location of the source is at the average elevation of the antennas. We verified that if the simulations are performed with the RFI sources located above the ground, the matched filter method recovers the source most strongly at the plane closest to the input height. However, the lack of antennas above the ground makes it more difficult to constrain the location of the RFI precisely in the vertical direction, and we see a significant contribution from the RFI source even at $Z=100\,$m. 

It is important to note that the 3D PSF of the image cube is strongly spatially dependent on the location of the source with respect to the array. The locations of RFI sources within the core are better constrained by the information in the visibilities, leading to sharper and more defined PSFs. For sources toward the edge of the array, the constraints are less strong in the radial direction, leading to radially extended PSFs. More specifically, the hyperbolic shape of the PSF for sources near the edge is because, given a baseline, the delay in the visibilities is the only information used by the matched filter method in constructing the image. Now, the set of possible locations for the RFI source, where the difference in its distances to the two stations equals the product of the delay value and $c$, forms a hyperbola with the two stations as its foci. This creates uncertainty in the location of the RFI sources along the hyperbola for a given baseline. The PSF captures the uncertainty in the location when such information from multiple baselines (i.e., multiple hyperbolas with different orientations) is combined. For RFI sources near the edge of the array, the hyperbolas corresponding to most baselines will point in the radial direction leading to radially extended PSFs, while for sources within the core, the location can be constrained better since the effects of radial extension per baseline is averaged out over many directions leading to a more symmetric PSF.

\subsubsection{Neccessity of the sequence of steps}\label{sec:matched_filter_alt}
The matched filter method used a specific sequence of steps described in Sect. \ref{sec:matched_filter_steps}, such as frequency averaging and computing the absolute value followed by baseline averaging. Omitting the absolute value step before baseline averaging leads to
\begin{align*}
\bigl<\langle \mathbf{V}_{pq}^{a}(\nu)\rangle_{\nu}\bigr>_{\mathrm{bl}} &= \langle P(\nu)\rangle_{\nu}\Biggl<\frac{\mathbf{E}_{pa} \mathbf{E}_{qa}^{H}G_{pqa}}{d_{pa}d_{qa}}\Biggr>_{\mathrm{bl}} \approx 0;
\bigl<\langle \mathbf{V}_{pq}^{b}(\nu)\rangle_{\nu}\bigr>_{\mathrm{bl}} \approx 0.
\end{align*}
Thus, even grid points where the RFI source is present have near-zero values due to the $G_{pqa}$ term, which can take both positive and negative values. Alternatively, omitting the frequency averaging step leads to
\begin{align*}
    \langle|\mathbf{V}_{pq}^{a}(\nu)|\rangle_{\mathrm{bl}} &= \Biggl<\frac{|P(\nu)\mathbf{E}_{pa} \mathbf{E}_{qa}^{H}|}{d_{pa}d_{qa}}\Biggr>_{\mathrm{bl}};
    \langle|\mathbf{V}_{pq}^{b}(\nu)|\rangle_{\mathrm{bl}} = \Biggl<\frac{|P(\nu)\mathbf{E}_{pb} \mathbf{E}_{qb}^{H}|}{d_{pb}d_{qb}}\Biggr>_{\mathrm{bl}}.
\end{align*}
This results in near-field maps with positive, nearly constant amplitude in all voxels. Both these effects were observed in matched filter near-field images made using the corresponding sequence of steps mentioned above.

The natural approach in matched filter imaging should be to omit both steps and just perform a baseline averaging after phasing. This is essentially the same as a Fourier transform operation used in far-field interferometric imaging. However, this results in
\begin{align*}
    \langle\mathbf{V}_{pq}^{a}(\nu)\rangle_{\mathrm{bl}} &= P(\nu)\Biggl<\frac{\mathbf{E}_{pa} \mathbf{E}_{qa}^{H}G_{pqa}}{d_{pa}d_{qa}}\Biggr>_{\mathrm{bl}}\approx 0,\nonumber\\
    \langle\mathbf{V}_{pq}^{b}(\nu)\rangle_{\mathrm{bl}} &= P(\nu)\Biggl<\frac{\mathbf{E}_{pb} \mathbf{E}_{qb}^{H} G_{pqb}}{d_{pb}d_{qb}}\times\mathrm{exp}\left[\frac{-2\pi \mathrm{i} \nu}{c}\Delta d_{pqab}\right]\Biggr>_{\mathrm{bl}}\approx 0.
\end{align*}
Here, we again get near zero values at both grid points containing RFI sources and those dominated by noise. We note that if the $G_{pqi}$ term was not present, only the contribution at grid point $b$ would approach zero (in the limiting case of a large number of non-redundant baselines). However, in the presence of baseline and source-dependent gains that are caused by using a coarse grid or because of not accounting for the array factor, this approach produces images with very low amplitudes. In Appendix \ref{sec:matched_filter_alt_data}, we demonstrate this effect on simulated and observed data.

\subsubsection{Limitations}
Although matched filtering is fast and enables quick identification of RFI-source locations, even for large numbers of visibilities taken over long integration times, there are two primary limitations of the method to provide comprehensive models of the identified RFI sources. Firstly, the matched filter operation does not correct for attenuation due to spherical wave propagation for each baseline-voxel pair and only ensures phase alignment\footnote{If the attenuation factor is artificially corrected for each voxel during imaging, the produced maps have high amplitudes near the edge of the image away from the stations where the distances between the voxels and all stations are very large.}. As a result, the maps cannot be converted to physical units necessary for building a model. It is worth noting that traditional far-field Fourier imaging is essentially a matched filter operation, but in that case, plane wave propagation does not involve amplitude attenuation, enabling the dirty image to be reconstructed in the physical units of the calibrated visibilities. Secondly, the matched filter implementation used in this study requires a frequency averaging step to down-weight contributions from voxels where RFI sources are not present. Ideally, this averaging should be performed over baselines after correcting for the expected near-field phase at each voxel. However, in practice, the physical extent of the stations limits the spatial resolution at which phases can be predicted, leading to phase errors that average out to zero over a large number of baselines. The effect of the phase errors is demonstrated in Appendix \ref{sec:matched_filter_alt_data} on simulated and observed data. To account for this, we included a baseline and source-dependent gain term in the formalism, which necessitates the frequency averaging approach instead. However, this results in a loss of spectral information, which is crucial for fully characterizing the RFI sources.

\begin{figure*}
    \includegraphics[width=2\columnwidth]{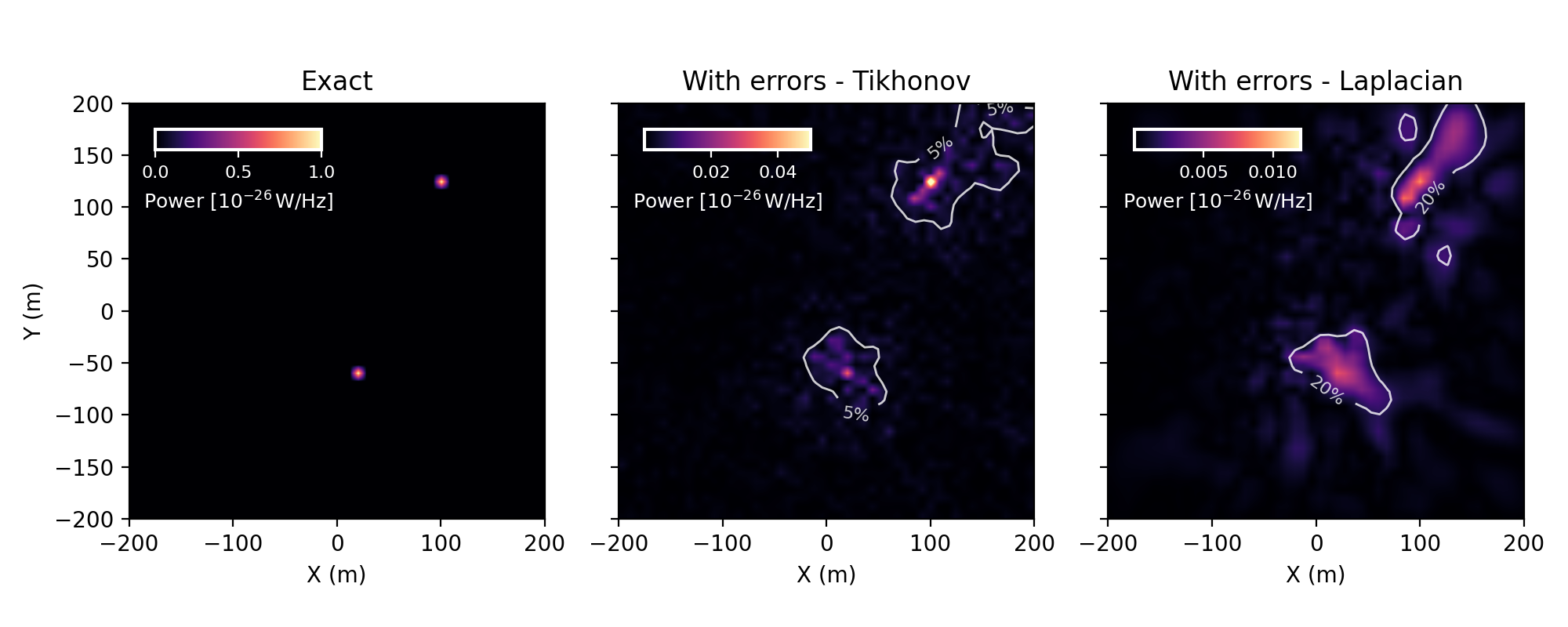}\\
    \includegraphics[width=2\columnwidth]{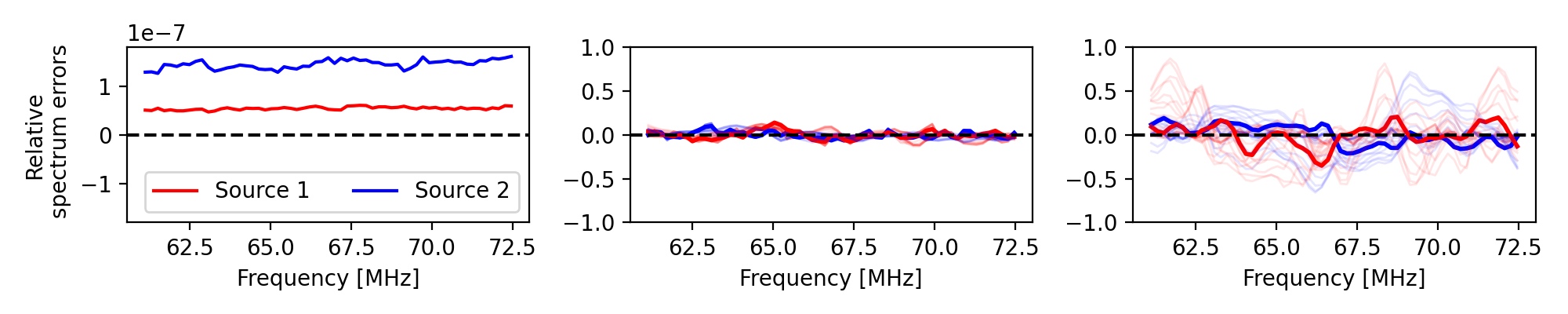}
    \caption{Images and spectra obtained from the MAP near-field imaging on simulated data. The leftmost column shows the exact case where the instrumental polarization is accounted for in the inversion. The next two columns show the inexact case corresponding to simulated data containing both off-grid and polarization-induced errors for Tikhonov and Laplacian regularization at $\rho=1.3\times 10^{-3}$ and $4.3\times 10^{-3}$, respectively. The top panels show the frequency-averaged images, and the bottom panels show the relative errors in the spectrum recovery for the three cases. The spectrum errors for the non-optimal $\rho$ values are shown in faded colors.}
    \label{fig:ls_reg}
\end{figure*}
\subsection{Maximum a posteriori imaging}
An alternative approach for near-field imaging from visibilities is to perform a maximum a posteriori (MAP) inversion of the near-field equation (Eq. \ref{eq:near_field}) to recover the spectral powers at a set of physical locations. Here, we ignore polarization effects, and the effect of this assumption is investigated later in this section. Similar to matched filter imaging, as a first step, the visibilities from Eq. (\ref{eq:near_field}) can be phased to the zenith in the far field by applying the inverse of the $W$ term, followed by time averaging to average out contributions from astronomical sources. The visibilities corresponding to a single element of the coherence matrix are then given by
\begin{equation}
V_{k}(\nu) = \sum_{i}M_{ki}(\nu)\times P_{i}(\nu), \textrm{where } M_{ki} = \frac{\mathrm{exp}\left[-\frac{2\pi \mathrm{i} \nu}{c}(d_{pi} - d_{qi})\right]}{{d_{pi} d_{qi}}}.
\end{equation}
Here, $k$ is a compound index of $pq$, which runs over all $N_{\mathrm{bl}}$ baselines.
The independent information from both the amplitude and phase can then be used to solve for the (real) power values by solving the system of equations for the real and imaginary parts together. This effectively recasts the equation into the form
\begin{align}
\begin{pmatrix}
\mathcal{R}[V_{1}(\nu)] \\
\vdots \\
\mathcal{R}[V_{N_{\mathrm{bl}}}(\nu)] \\
\mathcal{I}[V_{1}(\nu)] \\
\vdots \\
\mathcal{I}[V_{N_{\mathrm{bl}}}(\nu)]
\end{pmatrix}
&=
\begin{pmatrix}
\mathcal{R}[M_{00}(\nu)] & \cdots & \mathcal{R}[M_{0N_g}(\nu)] \\
\vdots & \ddots & \vdots \\
\mathcal{R}[M_{N_{\mathrm{bl}}0}(\nu)] & \cdots & \mathcal{R}[M_{N_{\mathrm{bl}}N_g}(\nu)] \\
\mathcal{I}[M_{00}(\nu)] & \cdots & \mathcal{I}[M_{0N_g}(\nu)] \\
\vdots & \ddots & \vdots \\
\mathcal{I}[M_{N_{\mathrm{bl}}0}(\nu)] & \cdots & \mathcal{I}[M_{N_{\mathrm{bl}}N_g}(\nu)]
\end{pmatrix}
\begin{pmatrix}
P_{1}(\nu) \\
\vdots \\
P_{N_g}(\nu)
\end{pmatrix},
\nonumber\\
\mathrm{or,}\ \mathbf{v}_{2N_{\mathrm{bl}}\times 1} &= \mathbf{M}_{2 N_{\mathrm{bl}} \times N_g}\mathbf{p}_{N_g \times 1}.
\end{align}
Here, $\mathcal{R}$ and $\mathcal{I}$ indicate real and imaginary parts, respectively. The system of equations is overconstrained if $2N_{\mathrm{bl}}>N_g$ and can be solved in a Bayesian framework. If instead $2N_{\mathrm{bl}}<N_g$, $N_{\mathrm{set}}$ frequency channels can be appended along the rows of the $\mathbf{v}$ and $\mathbf{M}$ matrices separately for the real and imaginary parts, under the assumption that the power is similar across these set of channels. The vector $\mathbf{v}$ and matrix $\mathbf{M}$ can then have dimensions of $2N_{\mathrm{bl}}N_{\mathrm{set}}\times 1$ and $2N_{\mathrm{bl}}N_{\mathrm{set}}\times N_g$ respectively, such that $2N_{\mathrm{bl}}N_{\mathrm{set}}>N_g$ is satisfied. Defining $\mathbf{n}_{N_g\times 1}$ as the instrumental noise vector, the linear system of equations for each set of frequency channels can be written as
\begin{equation}
    \mathbf{v} = \mathbf{M}\mathbf{p}+\mathbf{n}, \textrm{where }\mathbf{n}\sim\mathbf{N}(0,\mathbf{\Sigma}),
\end{equation}
where $\mathbf{\Sigma}$ is the noise covariance matrix, which we assume to be a diagonal matrix such that $\mathbf{\Sigma} = \sigma^2 \mathbf{I}$, $\sigma^2$ being the noise variance and $\mathbf{I}$ being the identity matrix. For unstable inversion problems where the matrix $\mathbf{M}^T\mathbf{M}$ is ill-conditioned, the stability can be improved by including a prior in the inversion using a regularization matrix $\mathbf{R}$. Assuming Gaussian likelihood and prior functions, the log-posterior is then given by
\begin{equation}
\log \mathcal{P}(\mathbf{p})=-\frac{1}{2}(\mathbf{v}-\mathbf{M} \mathbf{p})^T \mathbf{\Sigma}^{-1}(\mathbf{v}-\mathbf{M} \mathbf{p})-\frac{\rho^2}{2} (\mathbf{p}^T \mathbf{R}^{T}\mathbf{R} \mathbf{p})+\text { const.}
\end{equation}
Here, $\rho$ is a regularization parameter that controls the strength of regularization. The MAP estimate of $\mathbf{p}$, given by $\hat{\mathbf{p}}$, is equivalent to a regularized least squares solution for a diagonal noise covariance and is given by
{\small
\begin{align}\label{eq:two_norm}
\hat{\mathbf{p}}&=\left(\mathbf{M}^T \boldsymbol{\Sigma}^{-1} \mathbf{M}+\rho^2 \mathbf{R}^{T}\mathbf{R}\right)^{-1} \mathbf{M}^T \boldsymbol{\Sigma}^{-1} \mathbf{v}\: =\: \left(\mathbf{M}^T \mathbf{M}+\rho^2 \mathbf{R}^{T}\mathbf{R}\right)^{-1} \mathbf{M}^T \mathbf{v},\nonumber\\
&=\min_{\mathbf{p}}\left(\|\mathbf{v}-\mathbf{M p}\|^2+\rho^2 \|\mathbf{R}\mathbf{p}\|^2\right).\footnotemark
\end{align}
}
\footnotetext{Here $\sigma$ is absorbed into $\rho$.}
We used the \texttt{scipy.linalg.lstsq} function to estimate the power values \citep{virtanen2020scipy}. The routine employs the \texttt{gelsd} algorithm, which solves the least squares problem using singular value decomposition (SVD). In our case, the $\rho\mathbf{R}$ is appended to the matrix $\mathbf{M}$ along the rows, a zero vector is appended to $\mathbf{v}$, and the augmented system is solved using SVD, yielding the regularized solution.

\begin{figure*}
    \includegraphics[width=2\columnwidth]{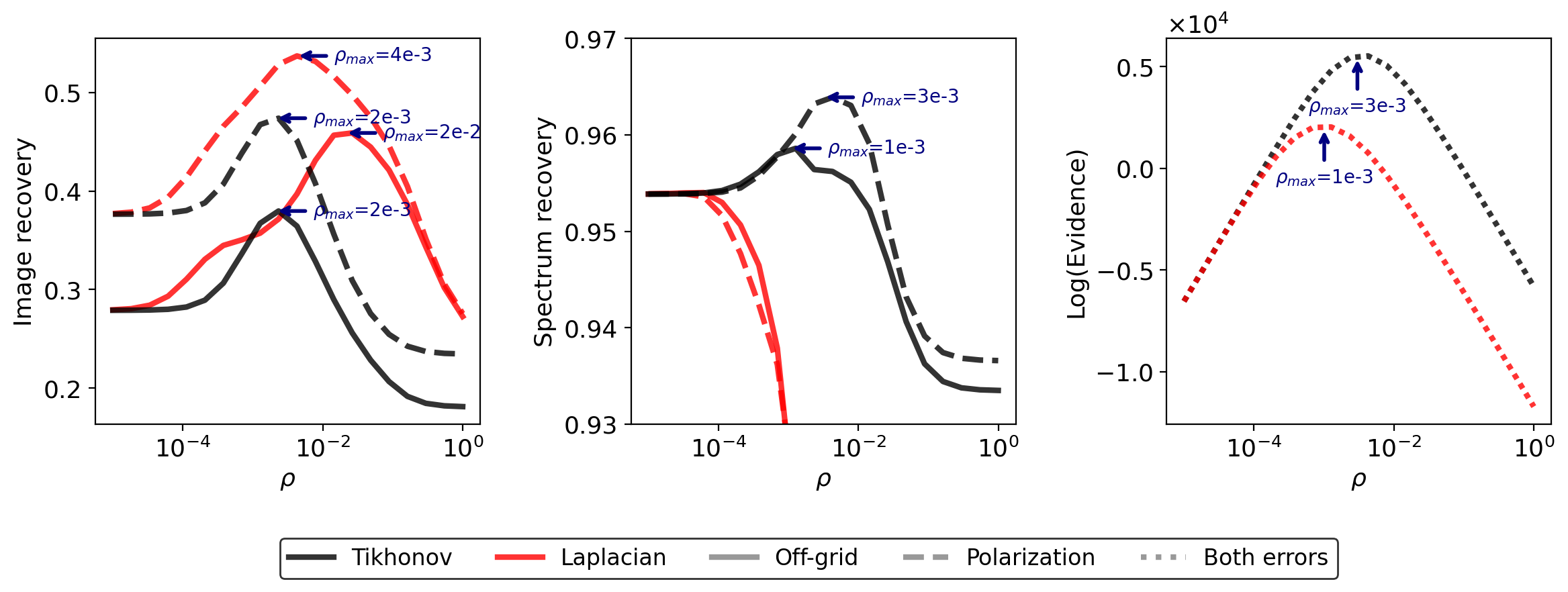}
    \caption{Optimal regularization parameters for image and spectrum recovery in MAP near-field imaging. The image recovery metric (left panel), the spectrum recovery metric (middle panel), and the log evidence (right panel) are shown as a function of $\rho$ for Tikhonov and Laplacian regularization for both off-grid and polarization-induced errors. The $\rho_{\mathrm{max}}$ corresponding to the peak values are indicated with blue arrows, except for the spectrum recovery metric for Laplacian regularization where a clear peak is not present.}
    \label{fig:ls_lam}
\end{figure*}
\subsubsection{Application to simulated data}\label{sec:ls_sim}
Similar to Sect. \ref{sec:mf_sim}, we used Eq. (\ref{eq:near_field}) to simulate visibilities for two RFI sources. The spectrum of the input RFI sources is assumed to be flat for simplicity, with a power of $10^{-26}\mathrm{W\,Hz}^{-1}$. We tested two types of regularization to stabilize the solution: Tikhonov and Laplacian regularization \citep{tikhonov1977solutions,belkin2003laplacian}. For Tikhonov regularization, $\mathbf{R} = \mathbf{I}$. Thus, it adds a penalty term proportional to the 2-norm of the solution vector and penalizes solutions deviating significantly from zero. Laplacian regularization penalizes differences between neighboring pixels in the solution, thereby encouraging spatial smoothness. For an $n$-dimensional spatial grid, we set $\mathbf{R} = \mathbf{L}$, where the Laplacian matrix $\mathbf{L}$ corresponding to the flattened solution vector containing $N_g$ grid points is given by
\begin{align}
\mathbf{L}_{i,i'} =
\begin{cases} 
+2n, & \text{if } i = i', \\
-1, & \text{if } i, i' \text{ are neighbors in the grid}, \\
0, & \text{otherwise}.
\end{cases}
\end{align}
Even without noise, there are errors introduced at two stages which makes the model deviate from the simulated $\mathbf{v}$ resulting in an inexact inversion problem, and least squares inversion becomes necessary to find the solution corresponding to the projection of the observed visibility vector in the range space of $\mathbf{M}$. The first deviation occurs when the simulated sources do not lie exactly on a grid point of the grid used in the image recovery. The second deviation arises when the visibilities are predicted using the instrumental polarization pattern imprinted into them, but the inversion is independently done per polarization component.  

We first tested the inversion for the exact case where the source locations in the simulated data lie exactly on two of the grid points used in the grid model during recovery. The instrumental polarization effect is accounted for by modifying $\mathbf{M}$ to include the product of the DD Jones matrices $\mathbf{E}_{pi}\mathbf{E}_{qi}^{H}$ for each ($2k,i$) index. The leftmost panel in Fig.~\ref{fig:ls_reg} shows the recovered image in this exact case (without regularization), and the two delta functions at the locations of the two input sources are recovered accurately. The bottom panel shows that the recovered spectra at the source locations have floating point errors of the order of $10^{-7}$. We note that this inclusion of the DD Jones matrices within $\mathbf{M}$ in the current configuration of a grid near the array cannot be done in data where there is a contribution from astronomical sources, since the included Jones matrices are not meaningful for the projected flux of far-field sources on the assumed grid. However, if the imaging is performed on a grid extending until $d_{\mathrm{F}}$, it is, in principle, possible to use DD Jones matrices that will reduce to those for astronomical sources in the far field limit.

For the inexact case, when the inversion is performed without regularization, the recovered images have large deviations from the input source distribution, with high values in pixels near the edge of the image. This makes it necessary to perform a regularized inversion. To find the optimal value ($\rho_{\mathrm{max}}$) where both the off-grid and polarization-induced errors are minimized, we tested the performance of Tikhonov and Laplacian regularization over a range of $\rho$ values. The MAP inversion was performed on a two-dimensional grid of $50\times 50$ grid points such that $2N_{\mathrm{bl}}>N_g$ is comfortably valid ($N_{\mathrm{bl}} = 3081$), and the inversion can be performed per frequency channel. We assume the grid to be located on the ground, with a resolution of $8\,$m. To assess the imaging accuracy as a function of $\rho$, we define a metric that calculates the normalized cross-correlation of the frequency-averaged recovered image against an input image that assumes a sinc response. We note that this is not the ideal metric since a sinc response would only be exact for a Fourier relation. A better metric for image recovery, based on the Bayesian evidence, is discussed later in this section. Spectra were estimated in a $16\times 16$ grid point box around each source. The spectrum recovery metric is defined as one minus the standard deviation of the fractional difference between the recovered and input spectrum. It should be noted that regularization often suppresses the values of the solution vector, because it acts as a prior on the solution vector that prefers values that are zero or constant\footnote{For Tikhonov regularization, the factor $\approx e_{i}/(e_i + \rho)$ where $e_i$ is the i-$th$ eigenvalue of $\mathbf{M}^T \mathbf{M}$.}. Thus, the overall factor is corrected for in spectrum estimation before computing the errors. For both Tikhonov and Laplacian regularization, we repeated the imaging for a range of $\rho$ values on visibilities corresponding to both off-grid errors, where the input source does not lie at a grid point of the chosen grid for recovery, and polarization errors, where the input visibilities contain the instrumental polarization effect but the inversion is performed from Stokes \textit{I} visibilities. We performed a similar exercise on Stokes \textit{V} data, which gave similar results. The results are shown in the left and middle panels of Fig.~\ref{fig:ls_lam}. The image recovery metric peaks close to a value of $\rho = 2\times 10^{-3}$ for both off-grid and polarization errors for Tikhonov regularization while for Laplacian regularization, the corresponding values are $\rho = 4\times 10^{-3}$ and $\rho = 2\times 10^{-2}$. However, in spectrum recovery, though Tikhonov regularization improves the metric in the $\rho$ range of $10^{-3}-10^{-2}$ where the fractional error is less than 4.5\%, Laplacian regularization significantly increases the errors beyond $\rho=10^{-4}$. Thus, since the regularization is performed spatially, the recovered image values for different frequencies can vary considerably, leading to large errors in estimated spectra.

Alternatively, we can estimate a Bayesian solution for $\rho$ itself by maximizing the evidence over a range of $\rho$ values. Following \cite{suyu2006bayesian} and \cite{ghosh2015bayesian}, the log evidence for the regularized solution can be written as
\begin{align}
\log P(\mathbf{v|\,\rho,\mathbf{R}}) =& -\rho^2 E_{S}(\mathbf{p_{\rho}})-E_{D}(\mathbf{p_{\rho}})-\frac{1}{2}\log(\det \mathbf{A})+N_g\log\rho \nonumber\\&-\frac{1}{2}\log(\det \mathbf{A}) -\frac{N_{\mathrm{bl}}}{2}\log(2\pi),
\end{align}
where $E_{S}(\mathbf{p}_{\rho})=||\mathbf{Rp}_{\rho}||^2$, $E_{D}(\mathbf{p}_{\rho})=||\mathbf{Mp}_{\rho}-\mathbf{v}||^2$, and $\mathbf{A} = \nabla\nabla(E_{D}+\rho^2 E_{S})$ is the Hessian matrix. We computed the log evidence values for both Laplacian and Tikhonov regularization. The results, shown in the right panel of Fig. \ref{fig:ls_lam}, account for both off-grid and polarization-induced errors in $\mathbf{v}$. We verified that when either off-grid errors or polarization errors were present, the results remained consistent. The log evidence peaks at similar values to those observed in the image and spectrum recovery metrics for both regularization methods. Tikhonov regularization has higher values of the evidence compared to Laplacian regularization over the entire range of $\rho$, indicating the suitability of Tikhonov regularization for this problem.

The frequency-averaged images for data containing both off-grid and polarization errors, corresponding to the $\rho=1.3\times 10^{-3}$ for Tikhonov and $\rho=4.3\times 10^{-3}$ for Laplacian regularization, are shown in the middle and right panels of Fig.~\ref{fig:ls_reg}. The contours at 5\% (for Tikhonov) and 20\% (for Laplacian) of the peak powers are indicated in white curves. While Laplacian regularization yields spatially smoother power distributions, the corresponding spectra have very large errors. We note that the matrix $\mathbf{R}$ corresponds to the inverse of the prior covariance matrix of the power distribution $\mathbf{p}$ \citep{vernardos2022very}. Since the input field is a set of delta functions or compact sources, their expected power spectrum is flat, and the corresponding two-point correlation function is a delta function. Hence, one would expect, a priori, that a Tikhonov regularization scheme would outperform those that prefer smooth solutions since the identity matrix is the inverse of a covariance matrix corresponding to a delta function correlation function.

\begin{figure}
    \includegraphics[width=\columnwidth]{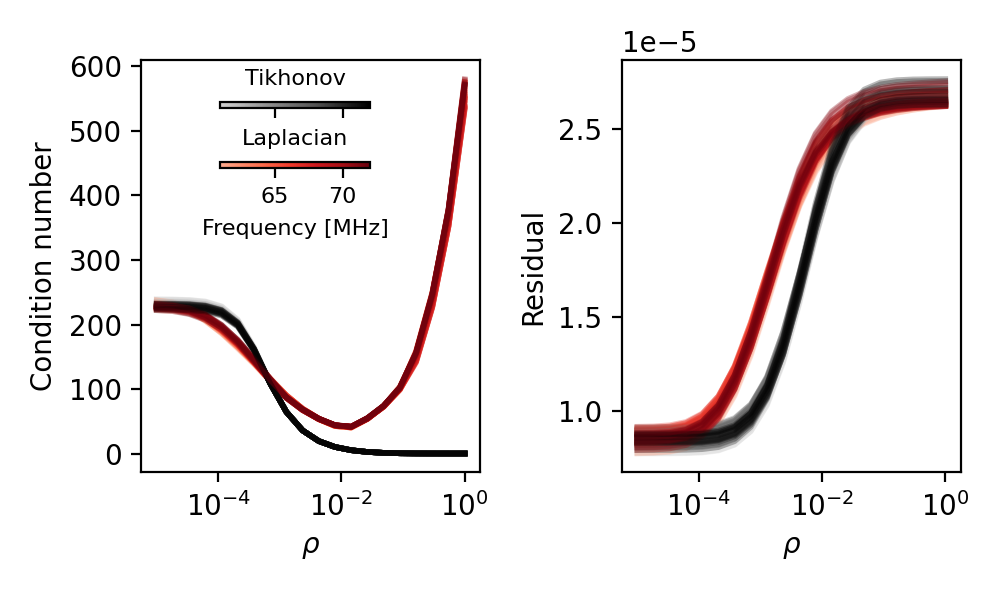}\vspace{-0.2cm}
    \includegraphics[width=\columnwidth]{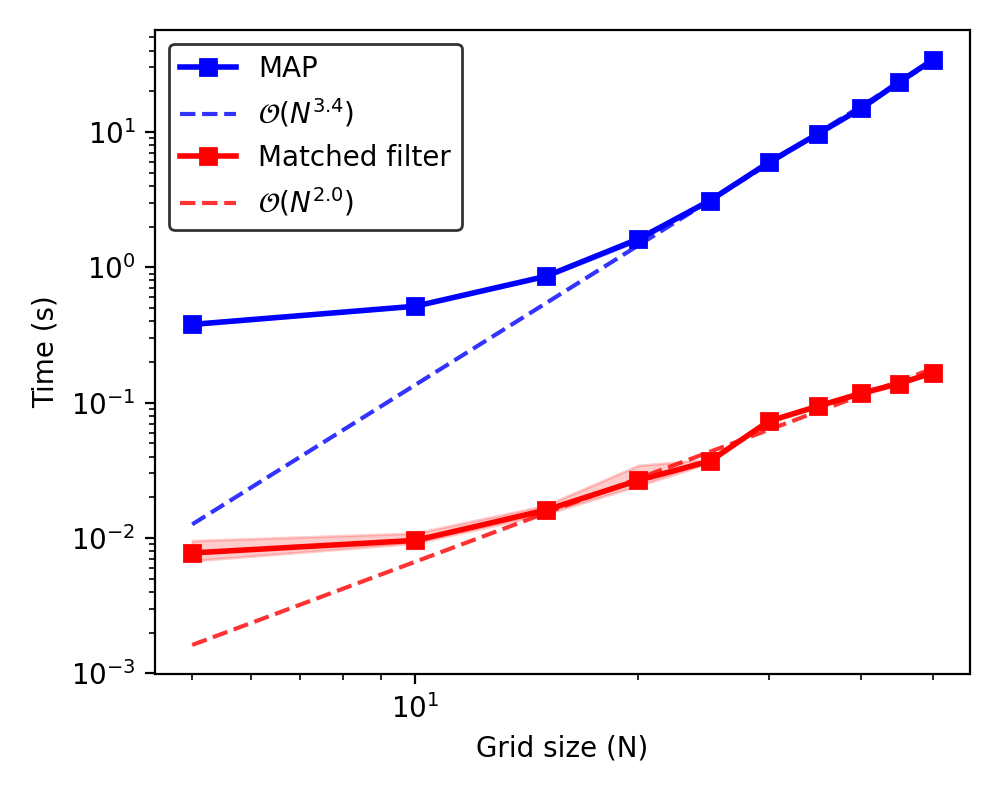}
    \caption{Stability and computational cost of the MAP and matched filter imaging algorithms. The top-left panel shows the condition number, and the top-right panel shows the least squares residuals from the MAP algorithm for both regularization schemes as a function of $\rho$ and frequency. The bottom panel illustrates the computational cost of the MAP and matched filter imaging algorithms. The dashed lines indicate the least squares fit to the last six data points in log-log space.}
    \label{fig:ls_mf_speed}
\end{figure}

The stability of the algorithm for both Tikhonov and Laplacian regularization as a function of $\rho$ is described by the condition number, which is the ratio of the largest to the smallest singular value in the SVD. This is calculated from the output of \texttt{scipy.linalg.lstsq} and is shown in the top-left panel of Fig.~\ref{fig:ls_mf_speed}. We find that while for Tikhonov regularization, the stability continues to improve with regularization, for Laplacian regularization, if we go beyond $\rho = 10^{-2}$, the stability starts decreasing as the condition number increases. The top-right panel in Fig.~\ref{fig:ls_mf_speed} shows the residuals in the least squares inversion given by the 2-norm in Eq. (\ref{eq:two_norm}) that is minimized. The residuals increase with the strength of regularization but are always higher for Laplacian compared to Tikhonov. The bottom panel of Fig.~\ref{fig:ls_mf_speed} compares the computational cost of the matched filter and MAP imaging methods as a function of grid size ($N=\sqrt{N_g}$) for a 2D square grid. We note that these results are for a single frequency channel, and multiple frequency channels could, in principle, be solved jointly given sufficient memory availability. The MAP method is slower by more than an order of magnitude compared to the matched filter algorithm. A straight line is fitted in log space to the last six data points in both curves to determine the computational cost at large $N$. The matched filter imaging has a computational cost of $\mathcal{O}(N^2)$, which is what we expect from performing the matched filter operation on $N^2$ grid points. The MAP inversion has a much higher computational cost of $\mathcal{O}(N^{3.4})$, making it computationally infeasible for large grid sizes. The two algorithms could be used in conjunction to build effective RFI models, with the matched filter method used to identify RFI sources and the MAP method used to perform detailed spatial and spectral characterizations on smaller grids constructed near the source locations (see Appendix \ref{sec:ls_joint}).
\subsubsection{Limitations}\label{sec:ls_limitations}
The current implementation of the MAP imaging has a few limitations. Firstly, although the inversion can be performed for the full polarization as demonstrated earlier, this is not feasible in actual data since the polarization effect that is included in the $\mathbf{M}$ during inversion is only correct for sources on the ground and power from sky sources will be scaled by incorrect factors. Thus, in actual data, the inversion needs to be performed per polarization, possibly in Stokes \textit{V}, where the data are less biased by emission from sky sources. An alternative approach would be to include far-field sources in the model and solve simultaneously for local RFI sources and astronomical sources, but this is left for future work. Secondly, the assumption that the shape of the recovered spectrum will not be affected by the polarization component on which the inversion is performed hinges on the fact that the spectrum shape measured by the dipoles does not depend on the direction of the incoming wavefront along the ground. This is only exactly correct for an infinitesimal dipole, and in practice, the shape of the spectrum will be dependent on the direction (see Appendix \ref{sec:dipole}). Still, for small bandwidth-to-central-frequency ratios where the dependence of the shape of the spectrum on the direction is relatively weak, the recovered spectrum shapes can be assumed to be constant across polarization, even for dipoles with a physical extent. There is always an overall scaling factor that is accrued, but that only depends on the geometry of the array and can be estimated through a separate cycle of simulation and recovery, as demonstrated in Sect. \ref{sec:far_field}. Thirdly, the systematic errors in the spectrum recovery are at $\approx 5\%$, which will limit the extent to which these spectra can be used to subtract the RFI power from visibilities through a direct prediction. We investigate this in more detail through simulations in Sect. \ref{sec:bias}.

\section{Local RFI sources near NenuFAR}\label{sec:rfi_nenufar}
In this section, we perform a detailed characterization of local RFI sources in the vicinity of NenuFAR from a subset of NenuFAR observations of the NCP field. The two local RFI sources identified by \citetalias{munshi2024first} were determined to be one of the causes of the strong excess variance seen in the data after foreground subtraction. The analysis by \citetalias{munshi2024first} adopted a simple approach of identifying and flagging entire baselines that were most affected by the RFI. However, the unflagged baselines continue to exhibit low-level RFI contamination and contribute to the excess variance, limiting the ability to constrain the 21 cm signal. In this section, we perform a detailed spatial, spectral, and temporal characterization of the local RFI sources in and around NenuFAR, using the algorithms developed in the previous section.
\begin{figure}
    \includegraphics[width=\columnwidth]{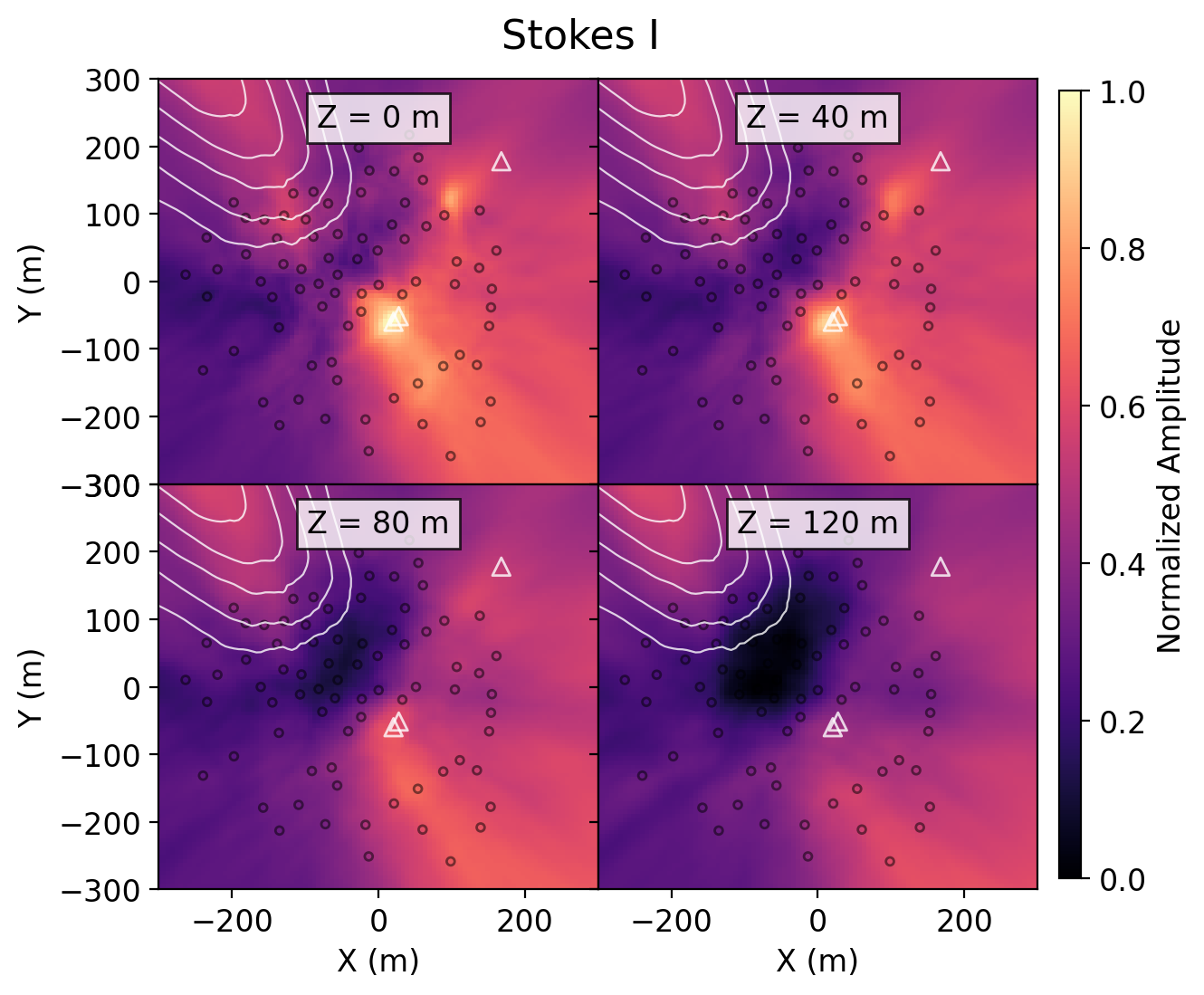}\\
    \includegraphics[width=\columnwidth]{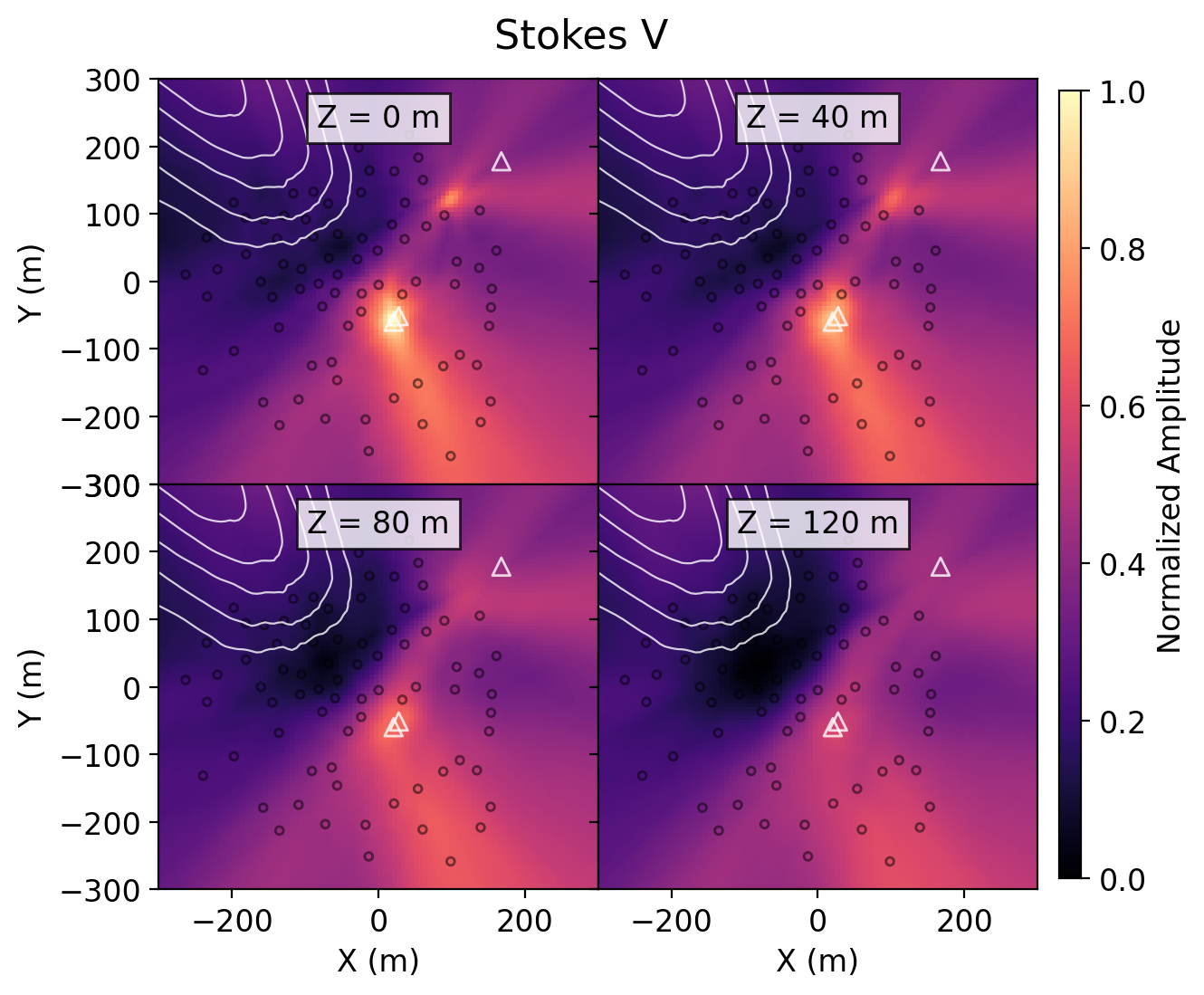}\\
    \includegraphics[width=\columnwidth]{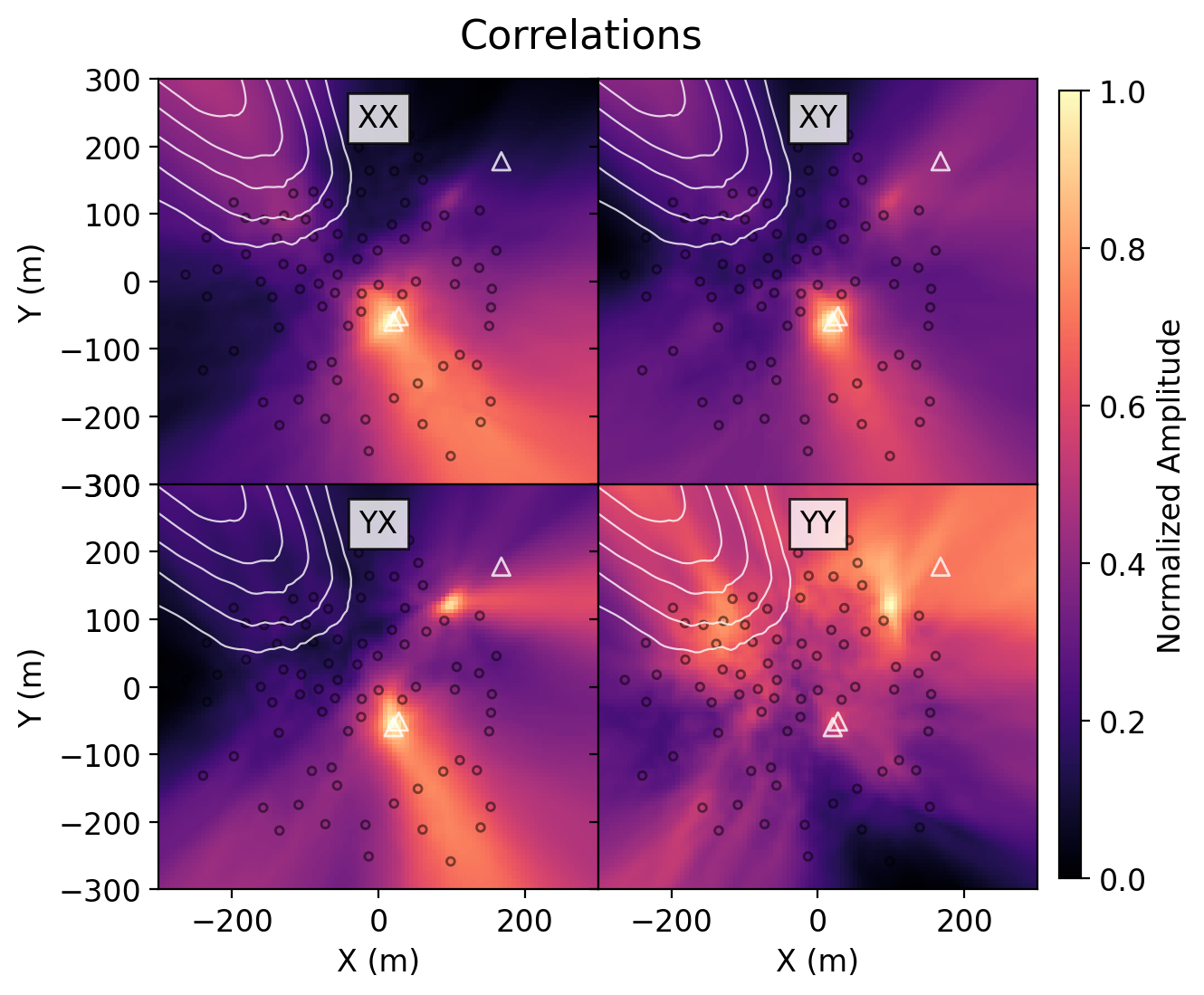}
    \caption{Matched filter images of NenuFAR data. Top panel: Stokes \textit{I} images at different heights ($Z$). Middle panel: Stokes \textit{V} images. Bottom panel: Images at $Z=0$ for different elements of the coherence matrix. Black circles represent NenuFAR core station locations, and triangles indicate buildings within the NenuFAR core. White contours indicate the projection of Cas A flux on the near-field domain.}
    \label{fig:nfi_mf}
\end{figure}
\subsection{Near-field imaging}\label{sec:nf_data}
To demonstrate the imaging techniques developed in the previous sections on actual NenuFAR observations, we used a 52 min subset of the same data as that used by \citetalias{munshi2024first}, which is less affected by the flux from strong A-team sources than other time intervals. The preprocessing and DD calibration-based sky-model subtraction performed to subtract Cygnus A, Cassiopeia A (Cas A), Taurus A, Virgo A, and sources in the NCP field above the confusion noise limit is described by \citetalias{munshi2024first}.
\subsubsection{Matched filter imaging}
We first applied the matched filter technique to NenuFAR observations of the NCP. The images were constructed on a $100\times 100\times 4$ grid with a grid resolution of $6\,$m in $X$ and $Y$ directions and $40\,$m in the $Z$ direction. In the top and middle panels of Fig.~\ref{fig:nfi_mf}, we present near-field images made from Stokes \textit{I} and Stokes \textit{V} data, respectively. The forward prediction of the visibility contribution of Cas A was imaged similarly, and the corresponding contours are overplotted in white lines indicating regions above 50\% to 90\% of the peak amplitudes, with five contour levels. Two local RFI sources can be identified to have the highest contribution at $z=0\,$m, which decreases as the height above the array increases. One of these RFI sources, as identified by \citetalias{munshi2024first}, is located in a building housing electronic containers. The localization is better in Stokes \textit{V} images, which are less affected by sky sources that have negligible intrinsic Stokes \textit{V} emission and only a low level of instrumental leakage from Stokes \textit{I} to \textit{V}. This is evident in the regions indicated by the expected contributions by Cas A, which have much lower amplitudes in Stokes \textit{V} compared to Stokes \textit{I}. Unlike local RFI sources, the amplitude of the projected power due to far-field sources does not decrease with height above the array. This is expected since the radiation received from these sources follows the plane wave approximation, and the matched filter operation will yield the same values at different heights. The bottom panel shows the images made per element of the visibility Jones matrix. Here, we see a similar signature to what was seen in images constructed from simulated data, including the instrumental polarization effect (left panel of Fig.~\ref{fig:mf_sim}). As we go from XX to YY, the amplitude of source 1 decreases while source 2 increases due to the orientation of the X and Y dipoles in NenuFAR and the locations of the RFI sources, as discussed in Sect. \ref{sec:mf_sim}. Thus, the data exhibit similar instrumental polarization signatures as predicted by the simulations performed with a simplistic $\sin\theta$ model for the dipole reception pattern.
\begin{figure}
	\includegraphics[width=\columnwidth]{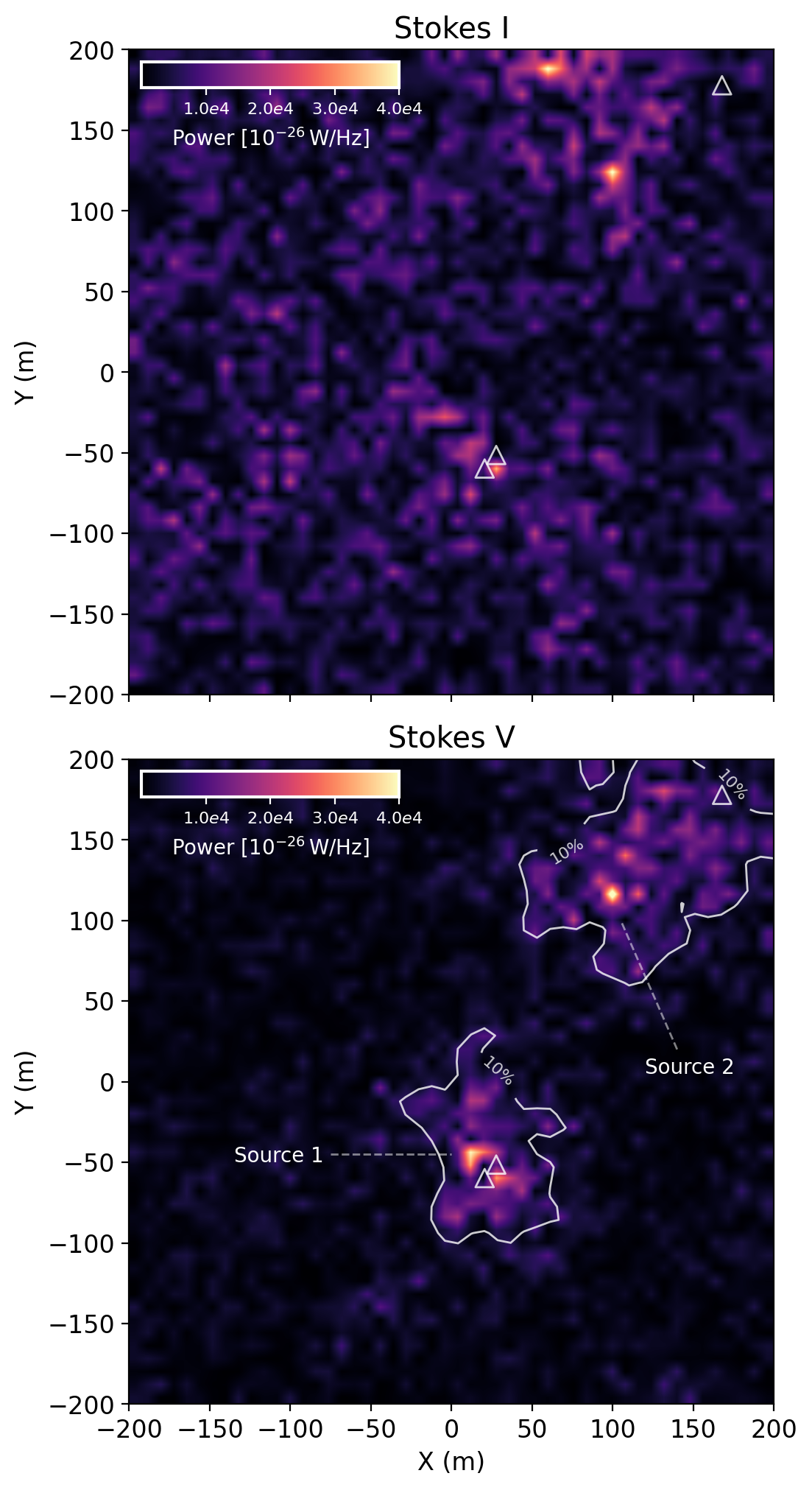}
    \caption{Near-field images of NenuFAR data made using the MAP method. The top and bottom panels correspond to frequency-averaged Stokes \textit{I} and Stokes \textit{V} images, respectively. The white curves in the Stokes \textit{V} image are contours indicating $10\%$ of the peak power level. The white hollow triangles indicate the locations of the buildings in NenuFAR.}
    \label{fig:nfi_ls}
\end{figure}
\subsubsection{MAP imaging}
The imaging was repeated using the MAP technique to retrieve the spectra and images in physical units. For all subsequent MAP inversions discussed in the paper, we used Tichonov regularization with $\rho=1.3\times 10^{-3}$. Fig.~\ref{fig:nfi_ls} shows the frequency-averaged images obtained using the MAP method. We find that the two sources are much more prominent in the Stokes \textit{V} image compared to the Stokes \textit{I} image since the inversion is less biased by sky sources in Stokes \textit{V}. A visual comparison against the top panel of Fig.~\ref{fig:nfi_mf} suggests that, in the presence of sky sources, matched filter imaging performs better than the MAP approach in recovering the spatial locations of RFI sources. The localization of the sources in Stokes \textit{V} is much better in the MAP approach than what was possible from the matched filter imaging. This is indicated in the Stokes \textit{V} image with white contours corresponding to the 10\% peak power level. This is possibly because images produced using the matched filter method are similar to dirty images, as discussed earlier, where the PSF blurs our view of the sources. Images made using the MAP method yield deconvolved images, where we are solving for the intrinsic distribution of power, enabling a sharper view of the sources. Here we see that the power from Cas A is not as clearly visible in the top-left part of the images as seen in matched filter images. This could be the case because we are solving for the sources in the near field in the MAP method instead of projecting the data into the near field. We also performed the imaging in smaller boxes with higher resolution around the two identified sources, which revealed that a joint inversion of the two sources is necessary to recover both the RFI source locations (Appendix \ref{sec:ls_joint}). Comparing the matched filter images in Fig.~\ref{fig:nfi_mf} with the MAP images in Fig.~\ref{fig:nfi_ls}, we see that while the matched filter images suggest source 1 to be significantly brighter than source 2, this is not the case in the MAP images. This possibly occurs because the matched filter imaging method does not account for the attenuation due to free space loss of spherical wave propagation, and since, on average, the stations are farther away from source 2, it infers a lower amplitude than the reality.
\begin{figure*}
    \includegraphics[width=2\columnwidth]{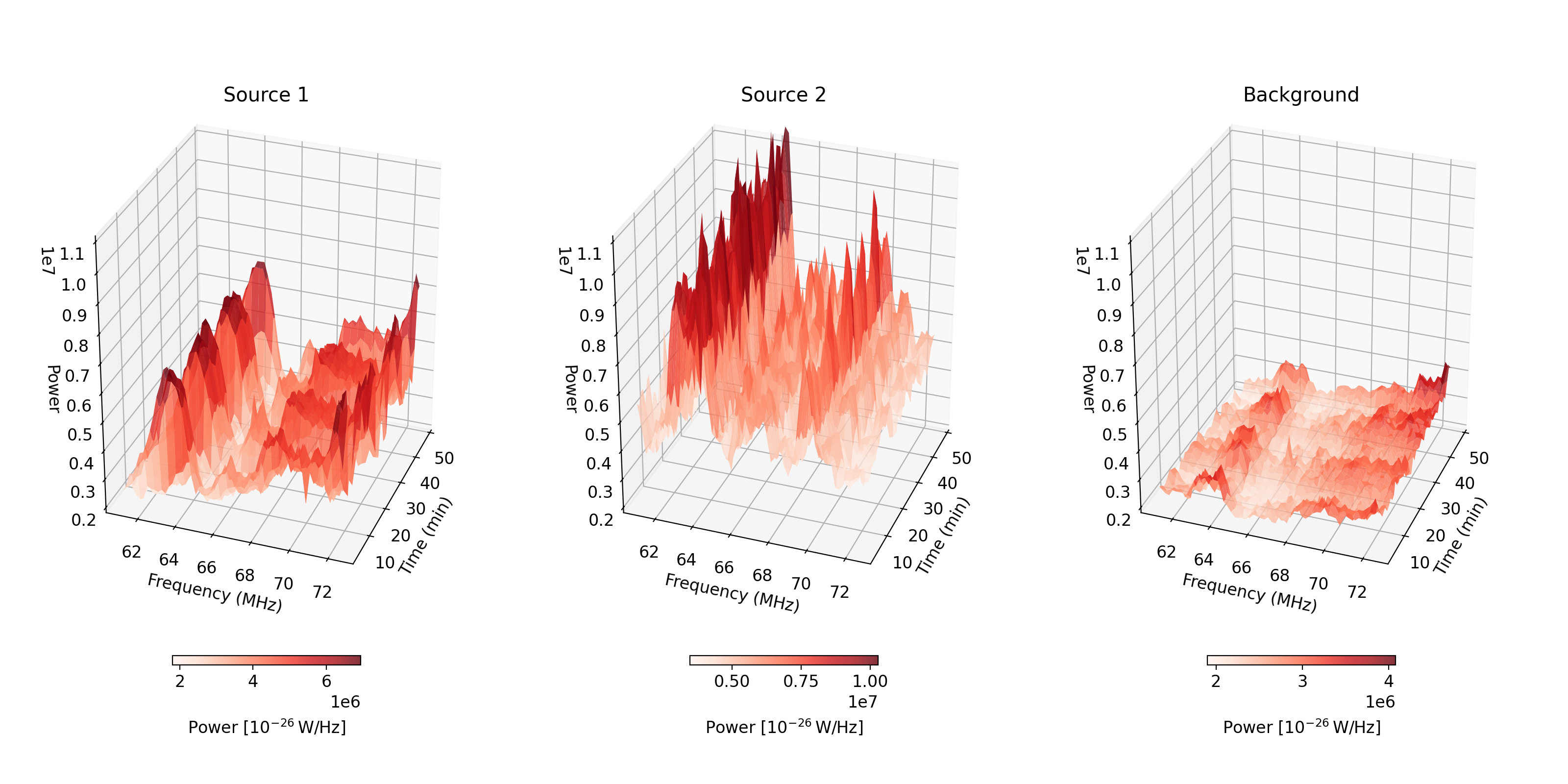}
    \caption{Spectral and temporal characterization from MAP near-field imaging. The left and middle panels show the recovered spectra of source 1 and source 2 as a function of time. The right panel shows the background spectra as a function of time.}
    \label{fig:rfi_wire}
\end{figure*}
\begin{figure}
    \includegraphics[width=\columnwidth]{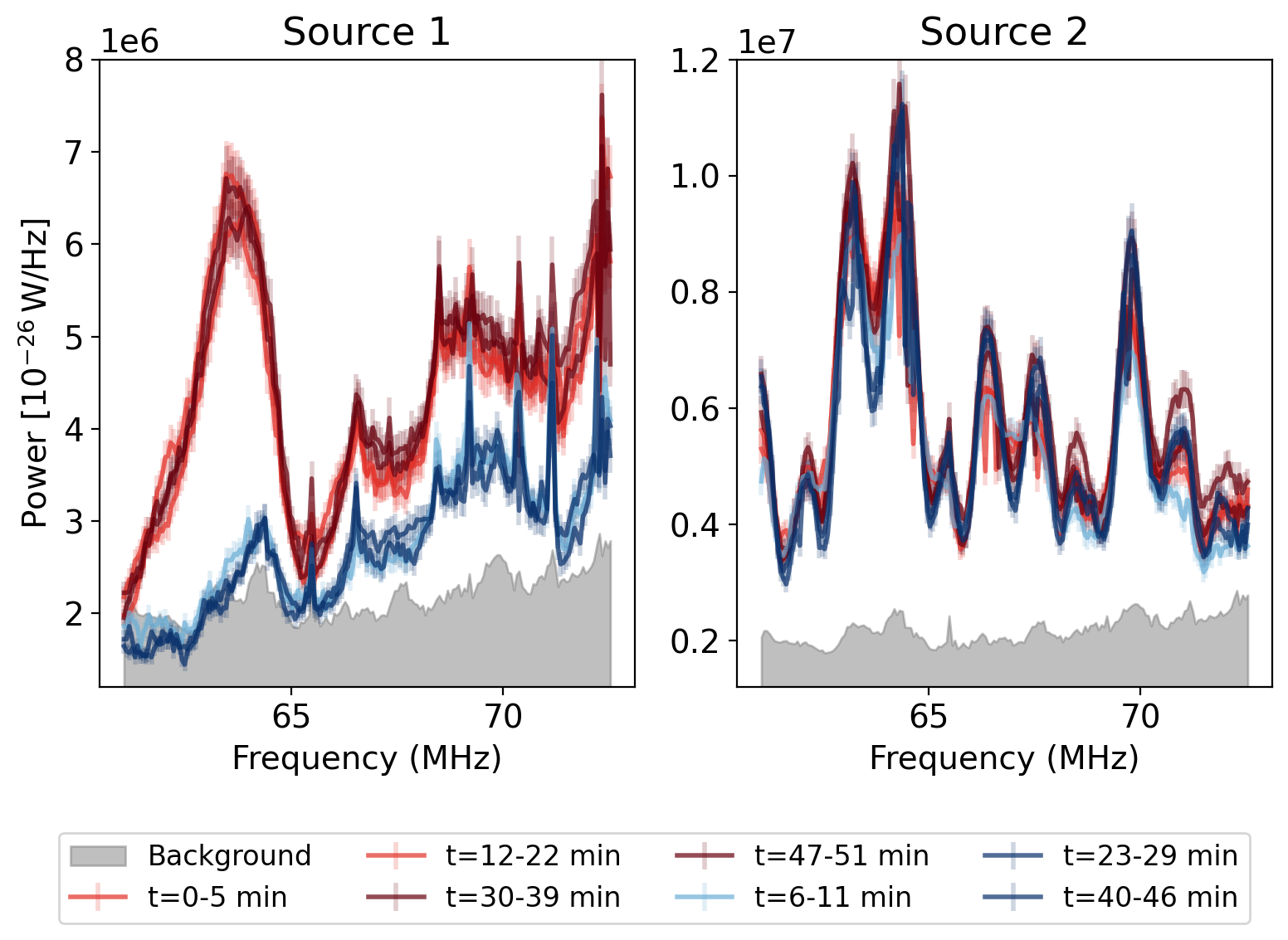}
    \caption{Spectra for the two sources estimated separately for different time segments. In both panels, the red lines show the spectra corresponding to the \textit{on} times of source 1 and the blue lines show the spectra for the \textit{off} times of source 1 identified from Fig. \ref{fig:rfi_wire}.}
    \label{fig:ls_spectra}
\end{figure}
\subsection{Temporal and spectral characterization}
Using the MAP method, we estimated the RFI spectra as a function of time. Images were constructed from Stokes \textit{V} visibilities at a 1 min time resolution. The spectra were estimated by summing up power values in $16\times16$ grid boxes containing the identified sources. We also estimated the spectrum in a $16\times 16$ box at the top left of the images where no sources are evident, as an estimate of the background power. The corresponding spectra as a function of time are shown in Fig.~\ref{fig:rfi_wire}. We find that source 1 has a periodicity in its intrinsic spectral power as a function of time and has distinct time intervals where it is bright or faint. Hereafter, we refer to these time intervals as \textit{on} and \textit{off}, respectively. For source 2, the power seems to increase gradually as a function of time. The periodicity of the source 1 amplitude was also observed by \citetalias{munshi2024first} in dynamic delay spectra constructed from visibilities. Also, source 2 was observed to increase in brightness compared to source 1 in matched filter near-field images made every 52 min interval of the 11.4 hour observation (not shown in the figure). The background should contain the noise power and any power spilled from the two sources due to the errors in the MAP reconstruction mentioned earlier and the incompleteness of the near-field model assumed. As indicated through the contours in Fig.~\ref{fig:nfi_ls}, power at a level of 10\% of the peak is confined to a region relatively close to the sources. However, we find that the background does have some leakage from the sources since it has higher values in the same frequency ranges as the source spectra, especially during the \textit{on} times. We constructed spectra for both sources separately using the data during each \textit{on} and \textit{off} time of source 1 identified in Fig.~\ref{fig:rfi_wire}. The corresponding spectra are shown in Fig.~\ref{fig:ls_spectra}. The spectral power is relatively consistent across the different time intervals, which are solved for independently. Source 1 clearly shows two different sets of spectra for the \textit{on} and \textit{off} times, while source 2 has similar spectral shapes across time. The possible origin of source 1 is air conditioning units in the buildings housing electronic containers, where we identified a hole in the Faraday cage used in shielding. The \textit{on} and \textit{off} times might thus correspond to the cooling cycles of the air conditioners. The identified location of source 2 corresponds to an inactive NenuFAR station, and it is currently not evident what the cause of the RFI signal is. While source 1 is seen in multiple nights of NenuFAR observations, source 2 is not as consistently present.

\section{Impact on far-field data products}\label{sec:far_field}
In this section, we assess the impact of local RFI sources on gridded data in the far field. We used the estimated RFI spectra from the 52 min of NenuFAR data, analyzed in the previous section, to simulate visibilities corresponding to the two local RFI sources using Eq. (\ref{eq:near_field}). For source 1, we used the background-subtracted \textit{on} spectrum as the input spectral power distribution to simulate visibilities only for the \textit{on} times. For source 2, we used the background-subtracted estimated RFI spectrum for the entire duration. A separate simulation and recovery were performed to estimate the scaling factor produced by the errors discussed in Sect. \ref{sec:ls_sim}, and the corrected spectrum was then used to simulate the full-resolution visibility data.
\begin{figure*}
    \includegraphics[width=2\columnwidth]{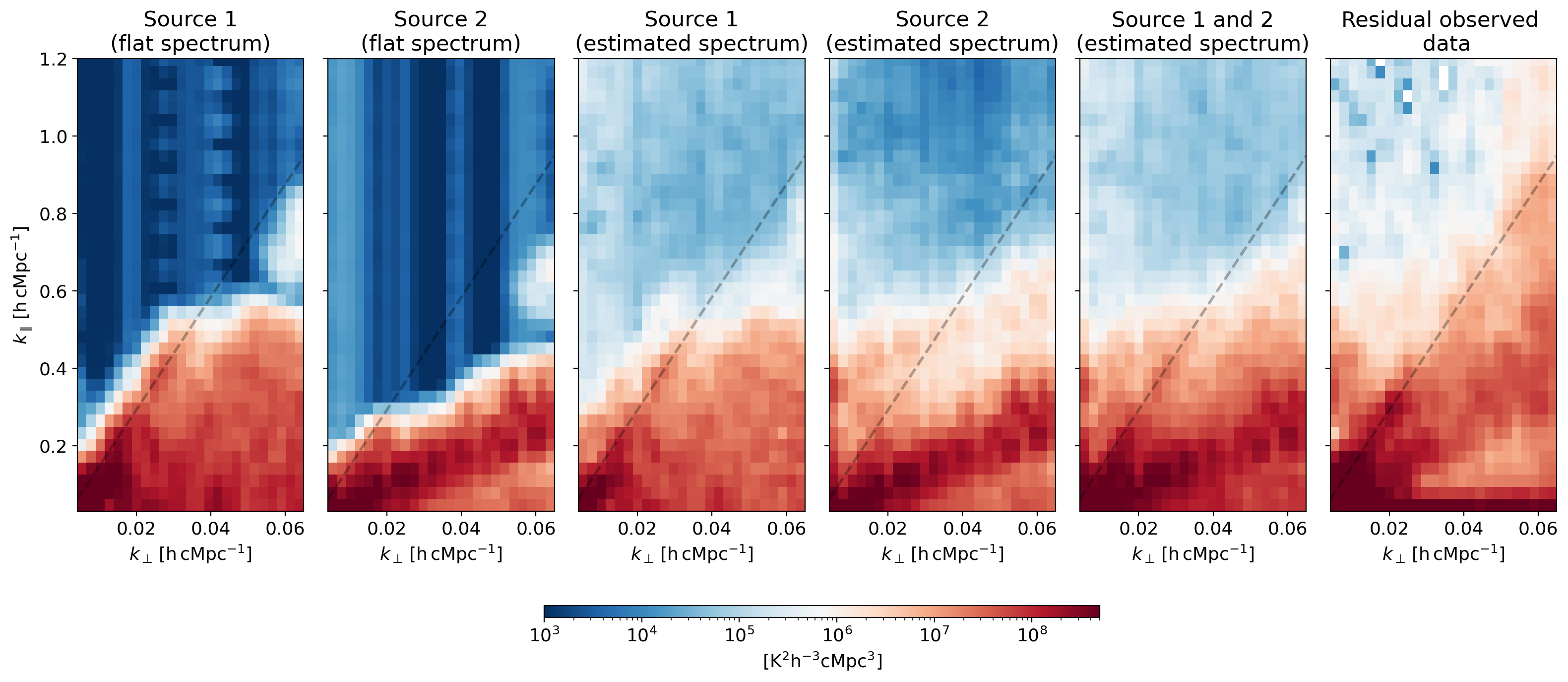}
    \caption{Impact of local RFI sources on the cylindrical power spectrum constructed toward the NCP field. The power spectra corresponding to a flat input spectrum are shown in the first two panels from the left. The next two panels show the power spectra corresponding to the estimated RFI spectra. The fifth panel from the left shows the total power from the two simulated RFI sources, and the right-most panel shows the residual data power spectrum after sky model subtraction. The black dashed lines indicate the horizon limit. We note that the noise power spectrum estimated from time-differenced Stokes \textit{V} data is subtracted from the residual data power spectrum to enable a one-to-one comparison.}
    \label{fig:ps_ncp}
\end{figure*}
\subsection{Impact on 21 cm power spectra toward the NCP field}\label{sec:impact_ncp}
The cylindrically averaged power spectrum is a commonly used metric in 21 cm cosmology, which shows the power as a function of spatial modes in the plane of the sky ($k_{\perp}$) and spatial modes along the line of sight ($k_{\parallel}$). To assess the impact of the two local RFI sources on the cylindrical power spectrum of the NCP field, we simulated the two sources separately, first with a flat spectrum with the average spectral power value across the band and then with the estimated RFI spectrum as discussed earlier. The power spectra were constructed using the \texttt{pspipe}\footnote{\url{https://gitlab.com/flomertens/pspipe}} routine, which uses \texttt{wsclean} \citep{offringa2014wsclean} to make image cubes, applies a Hann window in the image domain and Blackman-Harris filter along the frequency direction, and uses the modulus squared of the gridded data in the Fourier domain to estimate the power spectrum \citep{mertens2020improved}. Fig.~\ref{fig:ps_ncp} shows the estimated cylindrical power spectra. The black dashed lines indicate the full-sky horizon limit \citep{munshi2025beyond}, which defines the boundary of the range of modes that flat-spectrum foregrounds can occupy, known as the foreground wedge. The power spectra for the flat-spectrum RFI sources in the two leftmost panels show that the power is confined within the horizon limit. Source 1 fills up almost the entire wedge, while source 2 occupies primarily the lower $k_{\parallel}$ modes. This is because source 1 is within the core, and there is always a baseline for which source 1 is toward the physical horizon in the opposite direction of the phase center. Thus, following \cite{munshi2025beyond}, the condition for reaching the maximum horizon extent is met. So, RFI sources on the ground within the array will always fill up the entire foreground wedge, particularly for shorter baseline lengths where this situation of maximum delay is more likely to happen. We indeed note that for longer baselines, the power from source 1 does not always reach the horizon line. Source 2 is toward the edge of the core. As a result, the condition that a baseline points in the direction of the source and along the phase center is not necessarily met. Source 2 thus behaves more similar to a far-field source, and the modes it occupies depend on the location of the source and the phase center altitude, and it occupies a source wedge in the power spectrum. The boundary between the near and far fields in the context of delays and power spectra is demonstrated in Appendix \ref{sec:exp_delay}. The third and fourth panels from the left in Fig. \ref{fig:ps_ncp} show the power spectra for the simulations with the estimated RFI spectra, derived from those shown in Fig. \ref{fig:ls_spectra} as discussed above. The strongly fluctuating estimated RFI spectra result in leakage of power beyond the foreground wedge, leading to contamination of the EoR window, the region beyond the horizon limit. Source 2 power leaks well beyond the horizon line at the lowest $k_{\perp}$, possibly because of the strong fluctuations in the estimated source 2 spectrum. The fifth panel from the left shows the simulated power spectrum with both sources, and the rightmost panel shows the power spectrum constructed from the observed residual NenuFAR data for this 52 min segment after sky model subtraction. Comparing the simulated and observed data, it seems likely that the higher power along the horizon line and that at the very low ($k_\perp,k_{\parallel}$) values are caused by source 1, while the higher power at the lowest $k_{\perp}$ near $k_{\parallel}=0.35\,\mathrm{h}\,\mathrm{cMpc}^{-1}$ and that along the entire $k_{\perp}$ range at low $k_{\parallel}$ are likely caused by source 2. The data power spectrum has contributions from sky sources, which possibly causes some power to reach the horizon limit at high $k_{\perp}$. The higher power at the lowest $k_{\parallel}$ modes is due to the residual confusion noise limited power in the data after sky-model subtraction. These features are thus not present in the simulated RFI power spectra.

\subsection{Bias due to noise and 21 cm signal}\label{sec:bias}
The presence of thermal noise and 21 cm signal in the visibilities have the potential to bias the RFI spectrum estimation using the MAP approach. It is thus necessary to understand the nature of errors the algorithm induces, quantify the level of such errors, and identify the optimal approach for spectrum estimation to minimize them. In this section, we assess the robustness of the RFI spectrum estimation in the presence of noise and the 21 cm signal and its impact on the cylindrical power spectrum of the NCP field through forward simulations.

We generated 21 cm signal visibility cubes from the variational auto-encoder kernel trained on 21cmFAST simulations at a redshift of 20, as described by \citetalias{munshi2024first}. The simulated visibility cube was first converted to an image cube, and a simulated NenuFAR primary beam was applied, followed by the prediction of visibilities using the \texttt{predict} task in \texttt{wsclean}. The noise was simulated using NenuFAR's system equivalent flux density. For this high-noise scenario of a 52 min observation with NenuFAR, including signals at standard levels would have a negligible impact in addition to the noise. Assuming that, in practice, the spectrum estimation and subtraction would be performed per segment, we boosted the 21 cm signal fluctuations by a factor of 1000 to make it significantly stronger than the noise level for a single 52 min segment so that it has the potential to bias the RFI spectrum estimation.  The visibilities for the two RFI sources were simulated in the same manner as described at the beginning of Sect. \ref{sec:far_field}. The visibilities of the RFI, noise, and 21 cm signal were added to produce the simulated data set for this exercise.
\begin{figure}
    \includegraphics[width=\columnwidth]{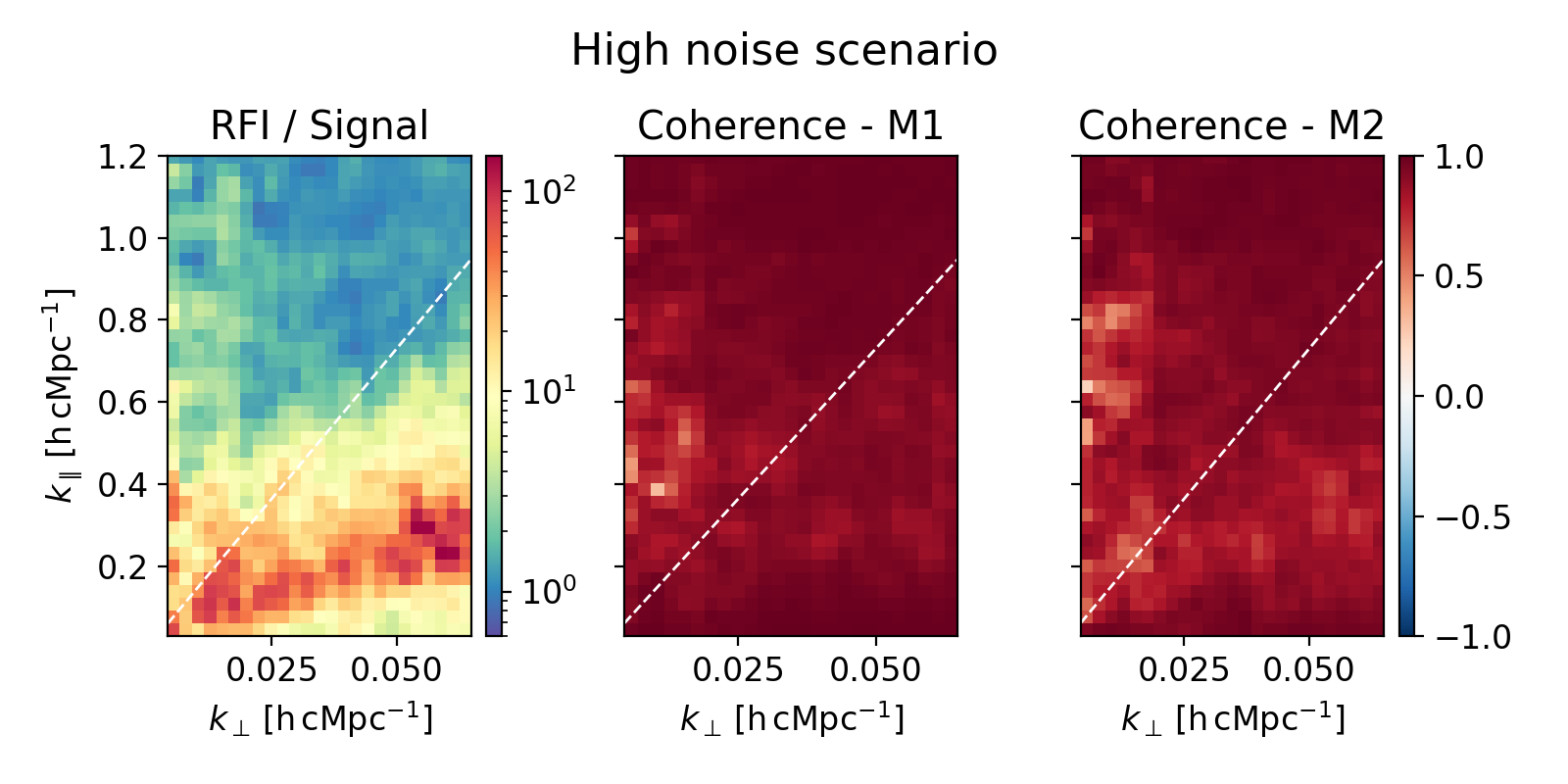}\\
    \includegraphics[width=\columnwidth]{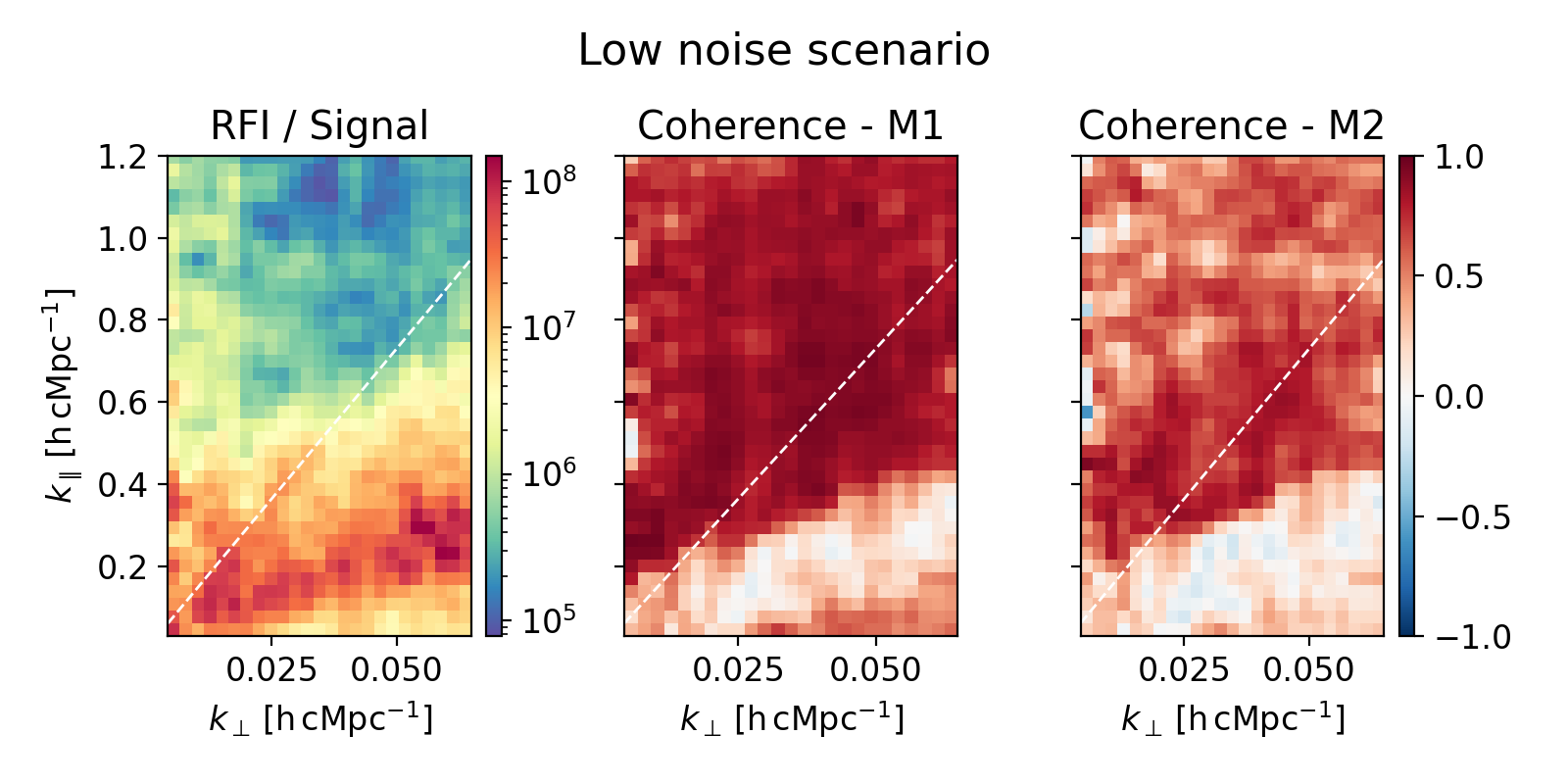}
    \caption{Results of robustness tests performed for the MAP spectrum estimation. The top row corresponds to the high noise and boosted signal scenario, while the bottom row corresponds to a low noise and standard signal scenario. The left column shows the dynamic range between the input RFI and 21 cm signal power spectrum in each scenario. The second and third columns show the cross-coherence between the noise-realization-subtracted residual Stokes \textit{I} data against the input signal corresponding to the two methods M1 (RFI spectrum estimation from Stokes \textit{V}) and M2 (RFI spectrum estimation from Stokes \textit{I}). The white dashed lines indicate the horizon limit.}
    \label{fig:sub_ps}
\end{figure}

Next, we estimated the RFI spectra from the \textit{on} and \textit{off} time segments of the simulated data separately, with a background spectrum estimated from the edge of the recovered image and subtracted from individual spectra. We note that, in addition to the RFI, the Stokes \textit{I} visibilities contain both noise and 21 cm signal, while the Stokes \textit{V} visibilities contain only noise. Thus, we use two methods: estimating the RFI spectra from Stokes \textit{V} (referred to as M1) and estimating them from Stokes \textit{I} (referred to as M2). Spectrum estimation in Stokes \textit{V} will be biased by noise, and that in Stokes \textit{I} will be biased by both the noise and 21 cm signal. The estimated RFI spectra were used to simulate visibilities, which were subtracted from the input Stokes \textit{I} visibilities to obtain the residuals. To understand the level of bias introduced in the RFI spectrum estimation by the presence of the 21 cm signal and noise, we use the cross-coherence metric (see Eq. (6) of \citealt{brackenhoff2024ionospheric}) between the residuals and the input 21 cm signal. To prevent decorrelation of the signal due to noise, we subtracted the noise realization cube from the residual data cube before computing the cross-coherence. The top row of Fig.~\ref{fig:sub_ps} summarizes the results of this exercise. The coherence is very well preserved for the boosted signal for M1, where the spectrum estimation is only biased by noise. For M2, the coherence is reduced in the modes where the RFI is strongest, indicating a slight bias in the spectrum estimation due to the presence of the signal. The top-left panel of Fig. \ref{fig:sub_ps} shows that the input RFI power is stronger than the boosted 21 cm signal by approximately two orders of magnitude at its peak, and the preservation of the coherence with the 21 cm signal indicates that the RFI spectrum estimation and subtraction performed in this exercise is able to reduce the RFI power by this factor. In the presence of systematic errors at a 5\% level induced by off-grid and instrumental polarization effects (as discussed in Sect. \ref{sec:ls_sim}), such a direct prediction and subtraction approach can, in principle, reduce the RFI power by up to a factor of 400. Thus, the coherence loss seen here is a combination of the bias due to noise and 21 cm signal as well as the systematic errors due to off-grid and polarization-induced errors.

To isolate the impact of noise and 21 cm signal on the RFI spectrum estimation and subtraction process, we next performed an ideal reconstruction with the source at a grid point and full polarization MAP inversion as discussed in Sect. \ref{sec:ls_sim}. This does not introduce systematic errors in spectrum estimation apart from floating point errors at a $10^{-7}$ level. For this exercise, we reduced the noise by a factor of 1000 and used standard 21 cm signal levels. The spectrum estimation and subtraction were performed in the same manner as before, and the cross-coherence between the noise-realization-subtracted residual data with the input 21 cm signal is shown in the bottom row of Fig.~\ref{fig:sub_ps}. Coherence to the signal is lost completely at the modes occupied by the peak RFI power, while in the EoR window, the coherence is largely retained for M1, and some coherence is lost for M2. We note that the dynamic range between the peak RFI power and the standard 21 cm signal is huge (bottom-left panel of Fig. \ref{fig:sub_ps}), and thus fractional errors introduced at a level of $10^{-4}$ or more in the RFI spectrum estimation due to the noise or 21 cm signal introduce a bias in the modes where the RFI power peaks. 

To test if these errors in M1 are actually caused by the thermal noise and are not systematic in nature, we repeated the spectrum estimation and subtraction from each time segment separately, as opposed to doing it separately for \textit{on} and \textit{off} segments. Not combining segments before spectrum estimation increases the residuals by a factor of $\approx 3$ in power (not shown in the figure), which is what we expect if the errors are due to noise in the data when 3 or 4 segments are averaged before the MAP inversion. Thus, decreasing the noise level would reduce this bias further for M1, and this was verified through a separate simulation. However, for M2, there will be some residual bias left due to the signal, and M1 offers a less biased way to estimate the RFI spectrum. Thus, estimating the RFI spectra from Stokes \textit{V} data, where it is less affected by astrophysical emission and the 21 cm signal, provides a more reliable approach.
\begin{figure}
    \includegraphics[width=\columnwidth]{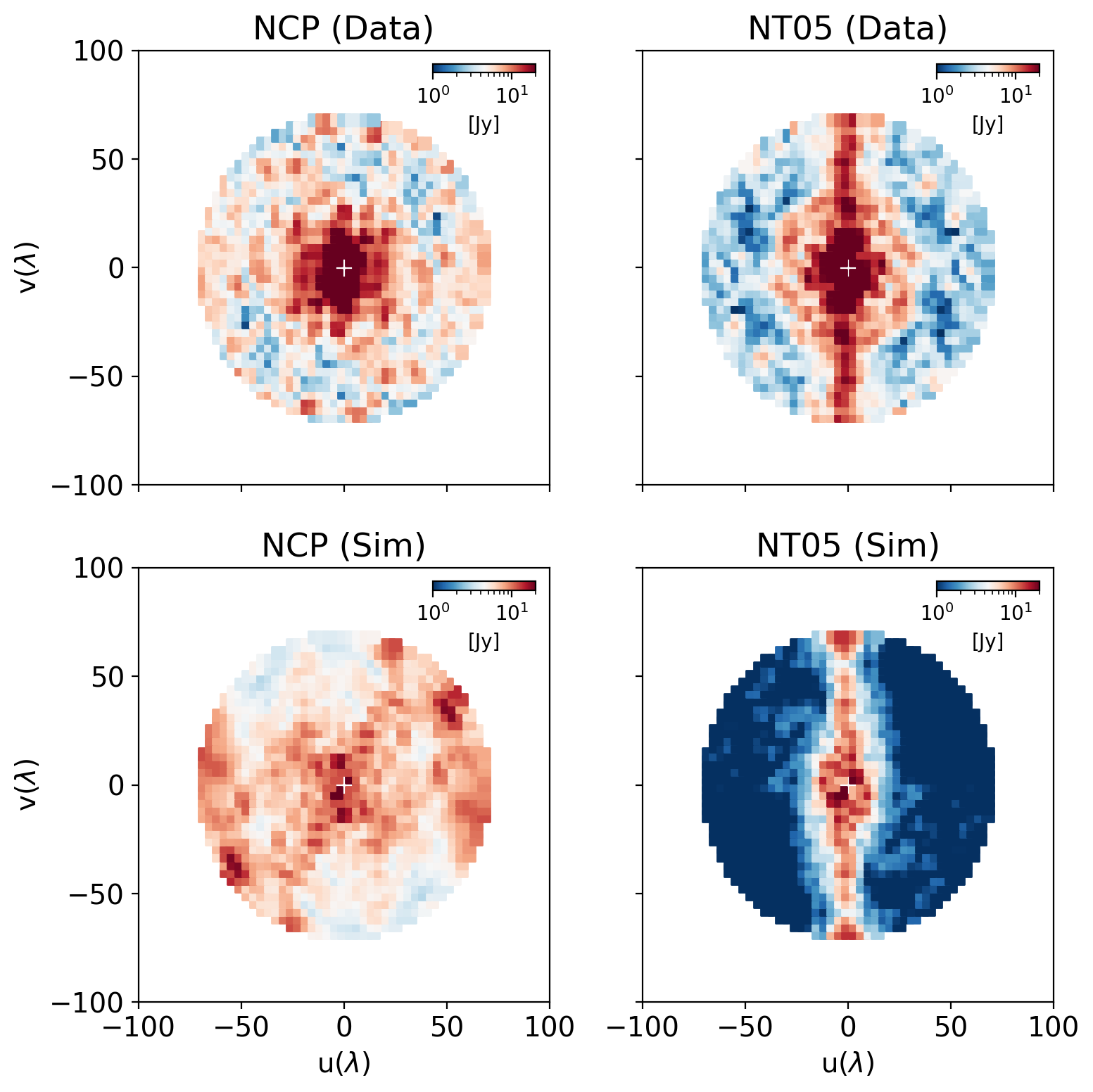}\\
    \includegraphics[width=\columnwidth]{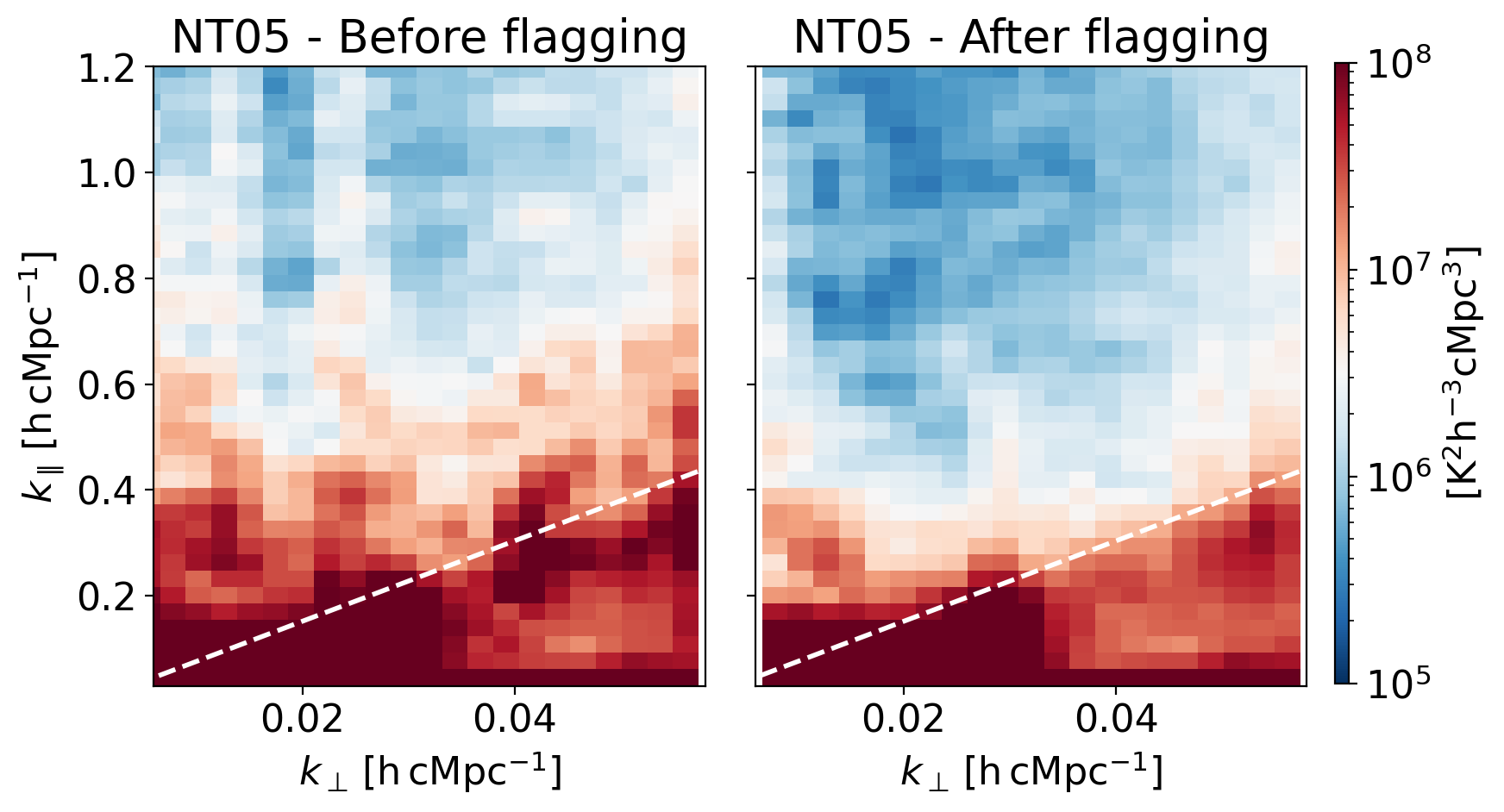}
    \caption{Impact of RFI sources on the $u\varv$ plane and power spectra of different target fields. The visibility amplitude in the gridded $u\varv$ plane for observed and simulated data for NCP and NT05 phase centers are shown in the top and middle rows. The bottom row shows the power spectra estimated from observed NT05 data, without (left) and with (right) the $u\varv$ cells around the $u=0$ line flagged. The white dashed lines indicate the horizon limit.}
    \label{fig:uv_fields}
\end{figure}
\begin{figure*}
    \includegraphics[width=2\columnwidth]{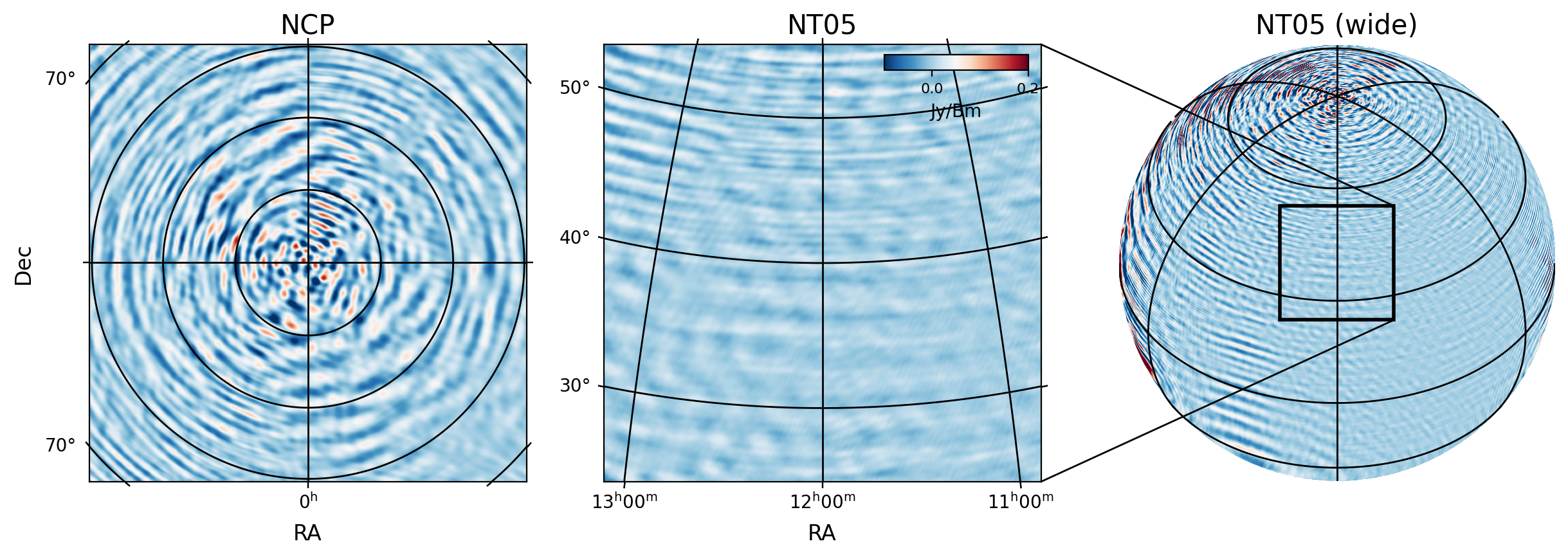}\\
    \includegraphics[width=2\columnwidth]{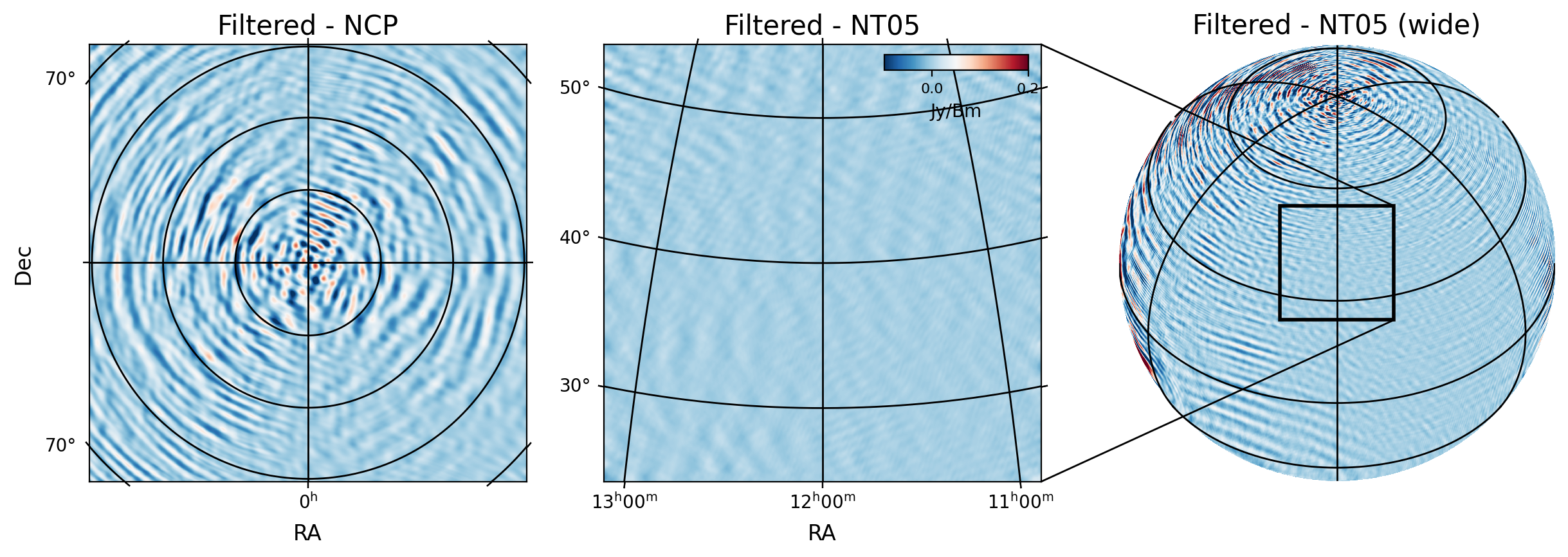}
    \caption{Impact of local RFI sources on far-field dirty images. The top row shows dirty images constructed from the simulated RFI visibilities, and the bottom row shows the images after the region  $|u| \leq 10$ is filtered out. The left and middle columns correspond to the NCP and NT05 fields, respectively. The rightmost column shows wide-field, full-sky images with the NT05 field indicated with a black square.}
    \label{fig:image_fields}
\end{figure*}
\subsection{Impact on observations of different target fields}
The NCP is a rather special point in the sky for synthesis observations since celestial poles are the only points that are stationary with respect to the array. For phase centers other than the NCP, gridding the data in the $u\varv$ domain fixed to the sky has the effect of averaging out the RFI as a function of time since the data are gridded into a domain where the sky sources add up coherently, and RFI, which are stationary on the ground, do not. However, in the special case of the NCP phase center, the RFI does add up coherently within the field of view (FoV) and behaves similarly to a source at the NCP. For small FoVs, the process of sampling by $u\varv$ tracks and gridding using a convolution kernel during the construction of the synthesis image results in a region of higher power along the source direction in the $u\varv$ plane (see \citealt{munshi2025beyond} for a mathematical treatment of this effect). For any phase center other than the NCP, the image is constructed to have the NCP toward the north. It thus follows that the power due to the RFI sources will have a peak along $u=0$ since the RFI imitates a source outside the FoV, toward the north of the image. A similar effect would happen for the south celestial pole in the southern hemisphere.

We performed simulations for the two RFI sources with the estimated RFI spectra during a 48 min observation of another target field with NenuFAR (NT05: RA=$12\,$h, Dec=$40^{\circ}$). The gridded data in the $u\varv$ plane and the power spectra were constructed in the same manner as was done for the NCP data and the simulations. The top and middle rows in Fig.~\ref{fig:uv_fields} show the frequency-averaged visibility amplitude in the $u\varv$ plane for both phase centers. In both the observational and simulated data, the power due to the RFI in the gridded $u\varv$ plane for the NT05 phase center is primarily confined to the region near $u=0$, while the RFI power is spread more across the $u\varv$ plane for the NCP phase center. We note that the actual data have noise and foregrounds while the simulated data do not, so only qualitative features can be compared between the observational and simulated data. The fact that the RFI does not add up coherently for the NT05 field is what results in its power being confined to a different set of modes than the sky power, enabling separation of part of the RFI power in this domain. The region around this $u=0$ line can now be flagged before power spectrum estimation to reduce contamination from the RFI sources. The bottom row of Fig.~\ref{fig:uv_fields} shows the power spectra estimated from observed data of the NT05 field before and after flagging five $u\varv$ cells in the vicinity of the $u=0$ line. This flagging step has a negligible impact on the NCP power spectrum (not shown in the figure), while the NT05 power levels are reduced significantly, especially in the EoR window.

Finally, we constructed images from the simulated data of the two RFI sources for both fields to illustrate the effect of local RFI in the image domain. The results are shown in Fig.~\ref{fig:image_fields}. For the NCP field, the RFI adds up at the center, producing PSF-like sidelobe ripples that resemble concentric circles around the center of the image. For the NT05 field, we see that the RFI-induced sidelobe ripples originate well outside the narrow field, toward the north of the image. The wide-field image illustrates that the origin of the sidelobes in the NT05 field is indeed the RFI power adding up at the NCP. Next, we filtered the region between $u=\pm 10$ for all three images. For the NCP field, filtering does not have a significant effect except for the portion of the sidelobes along the vertical direction whose amplitudes are reduced. For the NT05 field, the contamination from the RFI is significantly suppressed. Also, the reduction in contamination through filtering is lower for fields located close to the NCP. This is because for fields near the NCP, the effective RFI source is closer to the phase center, leading to slower phase ripples in the $u\varv$ plane, which are not suppressed as strongly by the convolution kernel, leading to wider signatures in the $u\varv$ plane. Thus, filtering in the $u\varv$ plane in a narrow region around $u=0$ suppresses a lower fraction of the RFI power.

\section{Summary and conclusions}\label{sec:discussion}
In this paper, we present and test two near-field imaging algorithms, matched filter imaging and MAP imaging, that can be used to identify and characterize local RFI sources from radio interferometric data. Both methods use the free space spherical wave equation and do not impose the plane wave approximation used in traditional interferometric far-field imaging. The two methods are demonstrated on simulated and observed data, and the impact of RFI sources on images, the $u\varv$ plane, and power spectra is studied through forward simulations. The main results from the paper are summarized below:

\begin{itemize}
    \item The matched filter imaging exhibits the signature of instrumental polarization in both simulated and observed data, which can be explained based on the location of the RFI sources and the NenuFAR array layout. Imaging on a 3D grid can be used to differentiate genuine local RFI sources from power due to far-field data projected onto the near-field domain. The RFI sources in NenuFAR are identified to lie very close to the ground. The matched filter method does not account for the attenuation due to spherical wave propagation, and hence, the produced maps do not have physical units. Errors in the estimated phase necessitate frequency averaging, resulting in images without spectral information, limiting the use of this algorithm to produce effective RFI models. Still, the matched filter imaging method is computationally fast, performs well in low S/N situations, and is robust to the presence of far-field sources, making it a useful technique for identifying the locations of local RFI sources.
    \item The MAP imaging performs a regularized inversion of the near-field visibility equation to produce maps in physical units with spectral information. Tikhonov regularization is found to give good results in both image and spectrum recovery on simulated local RFI visibilities. In observed data, the method performs well on Stokes \textit{V}, which is less affected by astronomical sources, and provides a sharper view of the RFI sources compared to matched filter imaging. We performed a spectral and temporal characterization of the two local RFI sources in NenuFAR data using the MAP imaging algorithm. One of the sources exhibits a periodicity in the intrinsic spectral power with distinct \textit{on} and \textit{off} time periods, while the other source exhibits a consistent spectrum with a gradual increase in average spectral power with time.
    \item Forward simulations of the RFI sources with a flat spectrum reveal that sources within the array fill up the entire foreground wedge in the cylindrical power spectrum toward the NCP field, while RFI sources outside the core occupy a source wedge similar to astronomical sources. Including the estimated RFI spectra in the forward simulations results in leakage of power into the EoR window and produces features that can be associated with corresponding features in observed residual NenuFAR data after sky model subtraction.
    \item We investigated the bias in the spectrum recovery in the presence of 21 cm signal and noise in simulated observations. Systematic errors due to off-grid and polarization errors limit the maximum fraction of RFI power that can be subtracted through a direct prediction at $\approx 95\%$. RFI spectrum estimation on Stokes \textit{V} data is biased only by noise, while that on Stokes \textit{I} data is biased by both noise and the 21 cm signal, making spectrum estimation from Stokes \textit{V} data a more reliable approach to minimizing bias.
    \item Finally, we analyzed the impact of RFI sources on far-field data products for different target fields through simulations. Since the celestial poles are stationary with respect to the array, the power due to RFI sources adds up at the poles. Thus, in far-field images of the NCP field, the RFI power is spread throughout the $u\varv$ plane. For other target fields far from the NCP, the RFI power in the $u\varv$ plane is confined to a narrow region around the $u=0$ line, enabling a reduction in RFI power in both images and power spectra by filtering this region out. Current NenuFAR observing programs thus focus on target fields besides the NCP to reduce the impact of these local RFI sources on the data.
\end{itemize}
Persistent low-level RFI that is not identifiable by traditional techniques in visibility dynamic spectra is one of the limiting factors for science cases such as 21 cm cosmology, which try to reach the instrumental thermal noise level to have the sensitivity to detect the faint 21-cm signal. These results show the potential of identifying such local RFI sources from radio interferometric visibilities and characterizing their temporal and spectral properties. This is an important step toward using these models to develop techniques to reduce the contribution of these local RFI sources from observed data. The simulation framework and near-field imaging algorithms are implemented in a Python library \texttt{nfis}\footnote{\url{https://github.com/satyapan/nfis}}, which is made publicly available. Improving these modeling techniques faces some critical challenges related to additional complexities, such as inadvertent reflections, cross coupling, and RFI emitter directivity. These effects need to be accounted for to develop more accurate models and techniques for efficient subtraction of the near-field RFI models from the data. This will be explored in future work. Although we have demonstrated these methods in the context of NenuFAR, these algorithms are applicable to any interferometer. Since the techniques operate on visibilities, temporarily storing them is necessary to model and subtract the RFI contribution, especially if the RFI spectrum varies with time. However, even in observatories where visibilities are not retained and are discarded immediately after processing, these methods can still be integrated into real-time workflows, serving as tools for generating diagnostic plots.
\begin{acknowledgements}
SM, LVEK, SAB, JKC, and SG acknowledge the financial support from the European Research Council (ERC) under the European Union’s Horizon 2020 research and innovation programme (Grant agreement No. 884760, "CoDEX”). FGM acknowledges the support of the PSL Fellowship. AB acknowledges financial support from the INAF initiative ``IAF Astronomy Fellowships in Italy'' (grant name MEGASKAT). EC would like to acknowledge support from the Centre for Data Science and Systems Complexity (DSSC), Faculty of Science and Engineering at the University of Groningen and from the Ministry of Universities and Research (MUR) through the PRIN project `Optimal inference from radio images of the epoch of reionization'. RG acknowledges support from SERB, DST Ramanujan Fellowship no. RJF/2022/000141. This paper is based on data obtained using the NenuFAR radiotelescope. NenuFAR has benefitted from the following funding sources : CNRS-INSU, Observatoire de Paris, Station de Radioastronomie de Nançay, Observatoire des Sciences de l'Univers de la Région Centre, Région Centre-Val de Loire, Université d'Orléans, DIM-ACAV and DIM-ACAV+ de la Région Ile de France, Agence Nationale de la Recherche. We acknowledge the Nançay Data Center resources used for data reduction and storage.
\end{acknowledgements}

{\renewcommand{\baselinestretch}{1} 
\small
\bibliographystyle{aa}
\bibliography{aa}
}

\begin{appendix}
\section{Matched filter imaging with baseline averaging}\label{sec:matched_filter_alt_data}
\begin{figure}[t]
    \includegraphics[width=\columnwidth]{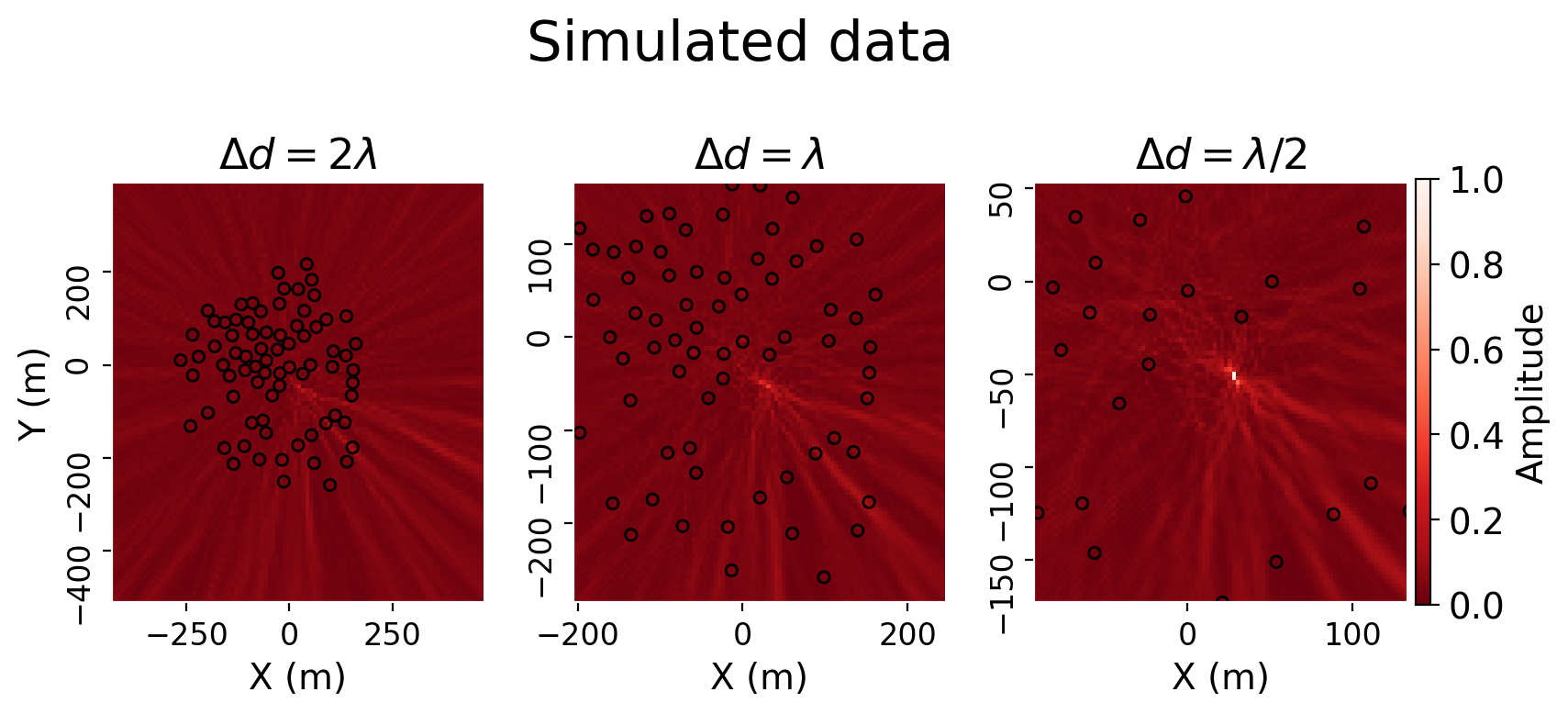}
    \includegraphics[width=\columnwidth]{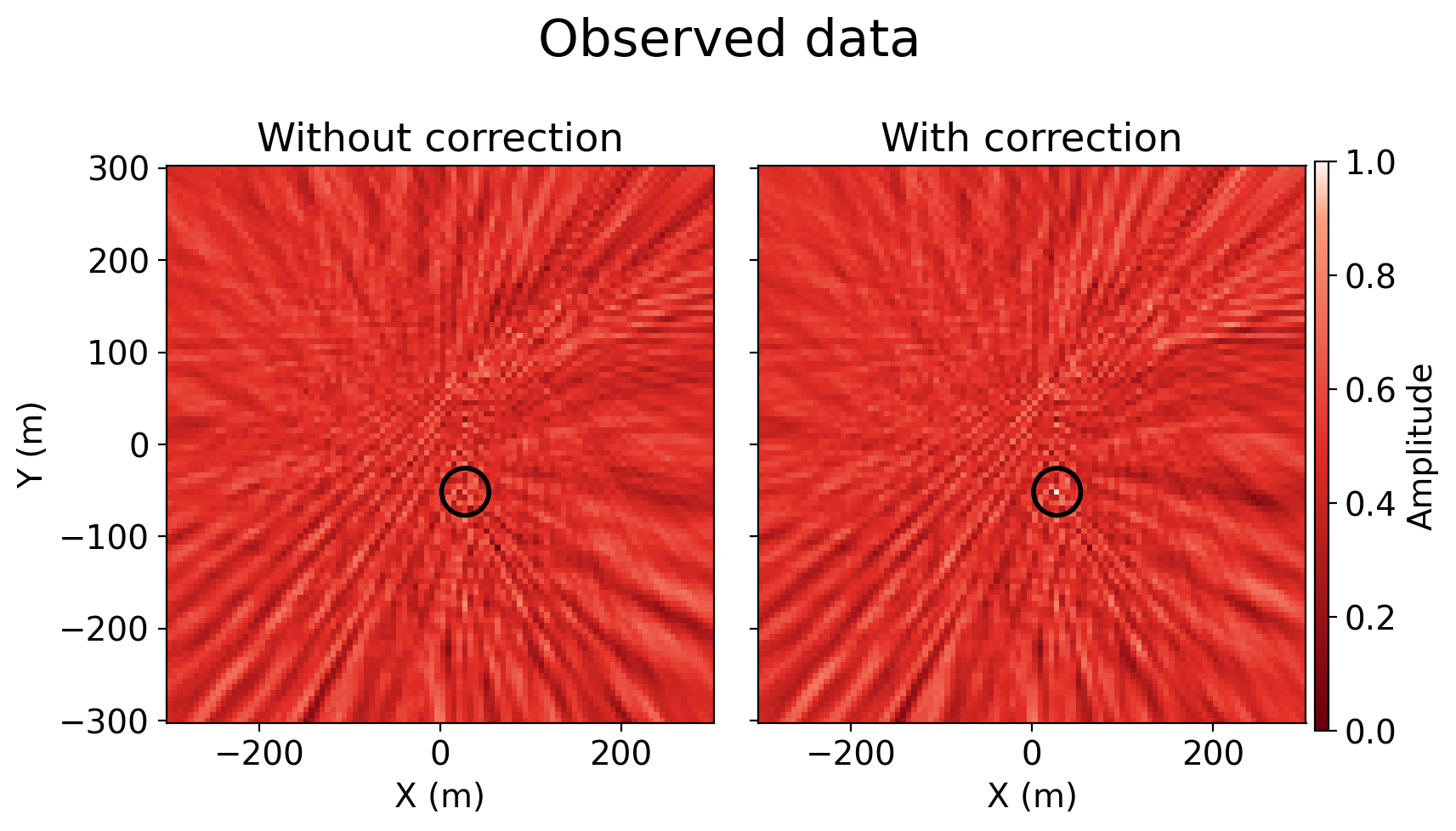}
    \caption{Matched filter near-field images with only baseline averaging. The three images in the top row (left to right) were constructed using progressively finer grid resolutions ($\Delta d$) from simulated data containing one local RFI source. Black circles indicate the NenuFAR station locations, providing a scale reference for the images. In the bottom row, the left panel shows the image made from Stokes \textit{V} observed data, and the right panel shows the image constructed after the signs of the baseline-dependent gains for source 1 have been estimated and corrected for. The location of the source is indicated with a black circle.}
    \label{fig:bl_gain}
\end{figure}
In Sect. \ref{sec:matched_filter_alt}, we discussed the effect of the baseline and source-dependent gain term that arises due to phase errors introduced by using a coarse grid or because of not accounting for the array factor. Here, we investigate this effect by simulating visibilities for a source located at the buildings within the NenuFAR core. Since the simulated visibilities do not account for the array factor, the uncertainty in phase due to the station extent is not an issue during the recovery, which assumes the stations to be point-like. Thus, we can, in principle, use arbitrarily high grid resolutions to make images using the baseline averaging approach. This is illustrated in the top row of Fig.~\ref{fig:bl_gain} by constructing matched filter images with the baseline averaging approach with different grid resolutions. Using a grid resolution of the order of $\lambda/2$ gives us a sharp view of the source, while a coarser grid essentially introduces $G_{pqi}$ due to phase errors and results in low amplitude images.

In actual data, $G_{pqi}$ exists intrinsically, even for the high-resolution grid, since the estimated phases used in matched filtering do not account for the locations of the antennas constituting a station. We constructed matched filter images using this approach on NenuFAR data. The bottom left panel of Fig.~\ref{fig:bl_gain} shows an example where we find an intensity distribution where the local RFI sources are not recovered. To illustrate the effect of the baseline and source-dependent gain term, for each baseline, we estimated the sign of the real and imaginary part of the peak value of visibility after applying the near-field phase corresponding to the location of source 1. Then, during the matched filter imaging, we corrected for the negative signs and constructed the image in the same manner. The result is shown in the bottom right panel of Fig.~\ref{fig:bl_gain}, where we can now recover the source. This exercise is only done for illustration purposes and cannot be done on actual data since the sign for each baseline will be different for each source. In that case, it would be necessary to predict the exact phase using information about the antenna locations and the pointing direction in the sky where the primary beam is steered.

\section{Impact of sources outside the grid in MAP imaging}\label{sec:ls_joint}
\begin{figure}[t]
    \includegraphics[width=\columnwidth]{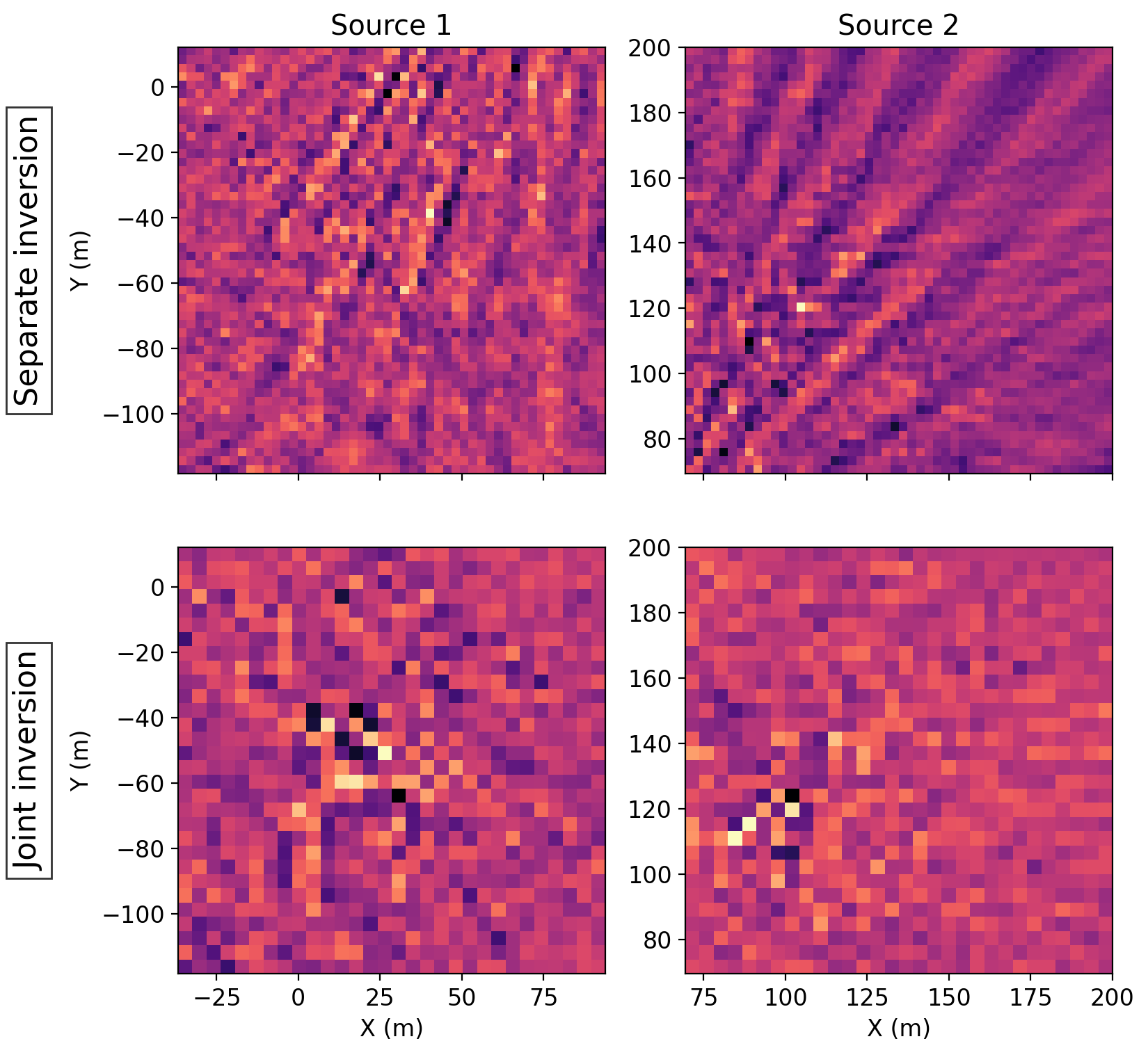}
    \caption{Near-field images made using the MAP method for small regions around the identified RFI sources. The top row corresponds to the images made separately for two grids. The bottom row shows the results of the joint inversion of two grids around the RFI source locations.}
    \label{fig:ls_joint}
\end{figure}
In MAP near-field imaging, it is important to make sure that the primary sources of near-field power lie within the grid used in the reconstruction. Here, we illustrate this on the same data as used in Sect. \ref{sec:rfi_nenufar}. First, the imaging was performed separately for two $50\times 50$ grids containing the identified source locations. The results are shown in the top row of Fig. \ref{fig:ls_joint}. We find that the point source distribution is not recovered, and in both images, we see fringes due to the power from the other source that is outside the grid. Next, the imaging was performed simultaneously for two $30\times 30$ grids containing the two sources. The grid size was reduced to make sure that the inversion problem remains sufficiently overconstrained per frequency channel since now pixels for both grids are being solved simultaneously. The results are shown in the bottom row of Fig. \ref{fig:ls_joint}. Here, the point source-like distribution is again recovered at the source locations identified in Fig. \ref{fig:nfi_ls}. We note that this requirement of both sources lying within the assumed grid is not necessary for matched filter imaging. This is again analogous to far-field imaging where the presence of the sky sources within the grid is not necessary to produce dirty images, but is essential to perform deconvolution, which is what the MAP approach to near-field imaging essentially does.

\section{Directivity and frequency dependence of dipole radiation}\label{sec:dipole}
In Sect. \ref{sec:ls_limitations}, we discussed that the recovered RFI spectrum is unaffected by the polarization from which it is retrieved, up to an amplitude factor. The radiation pattern of a dipole of length L is given by
\begin{equation}\label{eq:dipole}
R(\theta,\nu) = \frac{\cos\left(\frac{\pi\nu L}{c}\cos\theta\right)-\cos\left(\frac{\pi\nu L}{c}\right)}{\sin\theta}.
\end{equation}
Thus, $R(\theta_1,\nu)/R(\theta_2,\nu)$ is clearly frequency dependent, and the spectrum received by the dipoles from different directions is not the same up to a scaling factor. If we set $L\rightarrow 0$, we get
\begin{equation*}
R(\theta,\nu) \approx \frac{[1-\frac{1}{2}\left(\frac{\pi\nu L}{c}\cos\theta\right)^2]-[1-\frac{1}{2}\left(\frac{\pi\nu L}{c}\right)^2]}{\sin\theta}=\frac{1}{2}\left(\frac{\pi\nu L}{c}\right)^2\sin\theta.
\end{equation*}
Here, $R(\theta_1,\nu)/R(\theta_2,\nu)$ is frequency independent. Thus, the overall factor accrued in estimating the spectrum from Stokes \textit{V} data depends only on the location of the source compared to the array geometry, which can be accounted for through a separate simulation and recovery cycle as discussed in Sect. \ref{sec:far_field}. For small bandwidth to central frequency ratios, even for non-short dipoles, the frequency dependence of $R(\theta_1,\nu)/R(\theta_2,\nu)$ is weak, enabling spectrum estimation from a polarization component to good accuracy. For example, the spectral shape of the radiation pattern for $\theta=\pi/2$, calculated using Eq. (\ref{eq:dipole}) for a dipole of length $2.7\,$m at NenuFAR's frequencies, shows maximum deviations at the level of $8\%$. The maximum deviations occur in the case of extreme situations of the radiation coming along the axis of the dipole ($\theta=0$). In principle, if the exact radiation pattern is known, all these effects could be accounted for in the spectrum estimation.

\section{Near and far-field delays}\label{sec:exp_delay}
\begin{figure}[t]
    \includegraphics[width=\columnwidth]{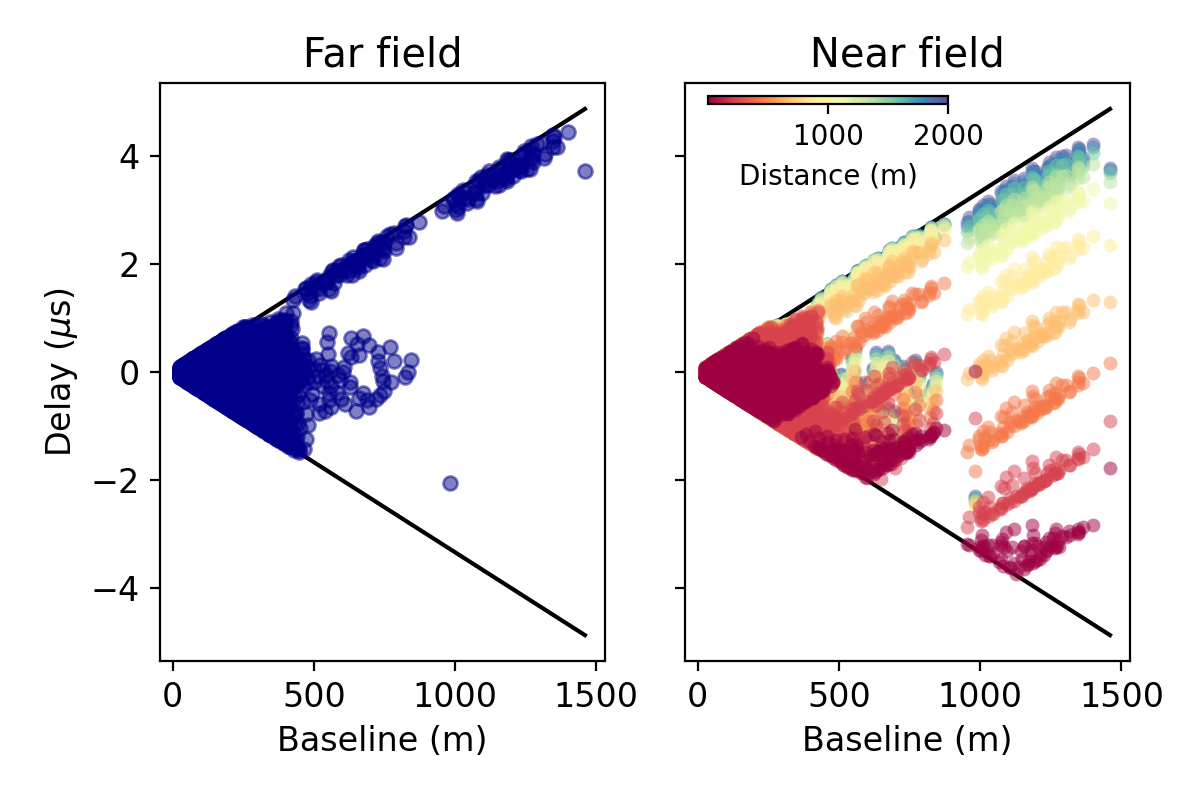}
    \caption{Expected delays due to a source located at different distances from the center of NenuFAR, calculated using the near-field and far-field equations. The solid black lines indicate the horizon delay limits.}
    \label{fig:exp_delay}
\end{figure}

In Sect. \ref{sec:impact_ncp}, we find that the RFI source located in the center of the core has a very different signature in the cylindrical power spectrum compared to the source located toward the northeast of the core, even when both of them have a flat spectrum. This effect can be demonstrated directly in the delay power spectra estimated from simulated observations with the RFI source placed at different distances from the center of the array. Instead of performing full visibility simulations for each RFI source location, we can directly estimate the expected delays based on the near-field and far-field phases under the assumption that the RFI spectrum is flat. We computed these delays corresponding to NenuFAR station locations for 10 different distances of the RFI source from the center of the array, with the source located due northeast. The results are shown in Fig \ref{fig:exp_delay}. The far-field delay power spectrum is unaffected by the distance of the source as long as the source vector remains in the same direction. The near-field delay power spectrum has very different signatures as the source is gradually taken from the center to the edge of the array and beyond. We note that as the source is gradually placed further from the center of the array, the longest baselines cross the boundary between the near and far fields last. The shorter baselines which are used to compute the cylindrical power spectrum already reach close to the far-field limit at a distance of $500\,$m, thus explaining why source 2, located at the edge of the $400\,$m core of NenuFAR, behaves as a far-field source would in the cylindrical power spectrum (Fig.~\ref{fig:ps_ncp}).

\end{appendix}
\end{document}